\documentclass{article}
\usepackage{arxiv}
\usepackage[utf8]{inputenc} 
\usepackage[T1]{fontenc}    
\usepackage{hyperref}       
\usepackage{multicol,multirow}
\usepackage{url}            
\usepackage{booktabs}       
\usepackage{amsmath,amssymb,amsfonts}
\usepackage{nicefrac}       
\usepackage{microtype}      
\usepackage{lipsum}		
\usepackage{graphicx}
\usepackage{natbib}
\usepackage{doi}

\newcommand{\vect}[1]{\boldsymbol{#1}}

\title{Identifying probabilistic weather regimes targeted to a local-scale impact variable}

\author{ Fiona R. Spuler \\
	Department of Meteorology\\
	University of Reading, Reading, UK\\
	\texttt{f.r.spuler@pgr.reading.ac.uk} \\
	\And
	Marlene Kretschmer \\
	Leipzig Institute for Meteorology\\
	University of Leipzig,\\
        Leipzig, Germany \\
 	\And
	Yevgeniya Kovalchuk \\
	Centre for Advanced Research Computing\\
	University College London\\
        London, UK \\
  	\And
	Magdalena Alonso Balmaseda \\
	European Centre for Medium-Range \\ Weather Forecasts\\
        Reading, UK \\
   	\And
	Theodore G. Shepherd \\
	Department of Meteorology,\\
        University of Reading \\
        Reading, UK \\
}

\renewcommand{\headeright}{Spuler et al. 2024}
\renewcommand{\undertitle}{}
\renewcommand{\shorttitle}{RMM-VAE}

\hypersetup{
pdftitle={RMM-VAE preprint},
pdfsubject={q-bio.NC, q-bio.QM},
pdfauthor={David S.~Hippocampus, Elias D.~Striatum},
pdfkeywords={weather regimes, variational autoencoders, Mediterranean precipitation, extreme events},
}
\begin{document}

\maketitle

Published in Environmental Data Science, doi \url{https://doi.org/10.1017/eds.2024.29.}\\
Received: 10 November 2023, Revised: 15 May 2024, Accepted: 09 August 2024 

\textbf{Abstract: }Large-scale atmospheric circulation patterns, so-called weather regimes, modulate the occurrence of extreme events such as heatwaves or extreme precipitation. In their role as mediators between long-range teleconnections and local impacts, weather regimes have demonstrated potential in improving long-term climate projections as well as sub-seasonal to seasonal forecasts. However, existing methods for identifying weather regimes are not specifically designed to capture the relevant physical processes responsible for variations of the impact variable in question. This paper introduces a novel probabilistic machine learning method, RMM-VAE, for identifying weather regimes targeted to a local-scale impact variable. Based on a variational autoencoder architecture, the method combines non-linear dimensionality reduction with a prediction task and probabilistic clustering in one coherent architecture. The new method is applied to identify circulation patterns over the Mediterranean region targeted to precipitation over Morocco and compared to three existing approaches: two established linear methods and another machine learning approach. The RMM-VAE method identifies regimes that are more predictive of the target variable compared to the two linear methods, both in terms of terciles and extremes in precipitation, while also improving the reconstruction of the input space. Further, the regimes identified by the RMM-VAE method are also more robust and persistent compared to the alternative machine learning method. The results demonstrate the potential benefit of the new method for use in various climate applications such as sub-seasonal forecasting, and illustrate the trade-offs involved in targeted clustering. \\
\\ 
\textbf{Impact Statement: }This paper introduces a new machine learning method for identifying large-scale atmospheric circulation patterns, so-called weather regimes, that modulate a local-scale impact variable such as extreme precipitation. This has the potential to enhance the usefulness of regimes for various climate applications such as impact-based sub-seasonal to seasonal forecasting or downscaling of climate model output. Co-authored by researchers with respective backgrounds in Meteorology and Computer Science, this paper is intended to introduce a new method in an accessible manner to researchers from both communities. Additionally, it aims to illustrate the similarities, differences and trade-offs associated with novel machine learning methods compared to more established statistical approaches for dimensionality reduction and clustering in weather and climate science.


\section{Introduction}\label{sec:intro}
\subsection{Weather regimes as mediators between local impacts and long-range teleconnections}

Large-scale atmospheric circulation modulates the occurrence of extreme events such as heavy precipitation and heat waves that cause devastating impacts to people and livelihoods across the planet. Understanding these dynamical drivers of local extreme impacts can both improve their near-term forecast skill for early-warning decisions \citep{coughlan_de_perez_adapting_2022, dunstone_windows_2023, gonzalez_weather_2022} and support the physical interpretation of future projected changes and associated uncertainty to identify robust adaptation pathways \citep{lemos_narrowing_2012, shepherd_storylines_2018}.

Weather regimes, defined as persistent and recurrent circulation patterns, are one common approach to understanding the low-frequency variability of the atmospheric circulation \citep{ghil_waves_2002, hannachi_low-frequency_2017, vautard_multiple_1990}. In many applications, weather regimes have proven particularly useful as discrete and interpretable mediators between long-range teleconnections in the climate system and local-scale impact variables \citep{beerli_stratospheric_2019, cassou_intraseasonal_2008, straus_preferred_2022, yiou_extreme_2004}. For example, over the North Atlantic and European region, weather regimes have been shown to carry a predictability signal from tropical teleconnections such as the Madden-Julian Oscillation and the El-Niño Southern Oscillation \citep{gadouali_link_2020, lee_enso_2019}, as well as from stratospheric polar vortex states \citep{charlton-perez_influence_2018, domeisen_role_2020}, while also modulating surface-level variables such as cold-extremes and precipitation \citep{ferranti_how_2018, pasquier_modulation_2019}. 

Due to these teleconnection signals as well as their persistence, weather regimes can improve both the skill and usability of forecasts for extended-range lead times \citep{allen_incorporating_2021, bloomfield_patternbased_2021}. On climate timescales, weather regimes have been used to disentangle the dynamic and thermodynamic components of climate change for extreme event attribution and quantify the role of atmospheric internal variability in observed trends \citep{cattiaux_winter_2010, horton_contribution_2015, terray_dynamical_2021}, as well as to statistically downscale climate models \citep{ailliot_spacetime_2009, maraun_precipitation_2010}. 

\subsection{Mediterranean weather regimes and precipitation over Morocco}\label{morocco_section}

The present study investigates the potential of targeted weather regimes to capture precipitation extremes over Morocco. The country is vulnerable to both flooding driven by extreme rainfall, which has caused over 760 million USD in economic damages since 1950 \citep{delforge_em-dat_2023}, as well as droughts which have threatened food security, agricultural livelihoods and compounded debt crises of the country \citep{tanarhte_severe_2024}. Extreme precipitation events primarily occur in extended winter, between November and March, and can lead to different types of flooding events, ranging from gradual or flash floods of wadis (river valleys), to torrential flash floods of small mountain basins or flooding of urban areas \citep{loudyi_flood_2022}.

Previous literature has investigated the dynamical drivers and precursors of wintertime extreme precipitation over the Western Mediterranean region, highlighting dynamically driven moisture flux from the Atlantic as a key driver \citep{dayan_review_2015, ulbrich_climate_2012} and positive anomalies in potential vorticity over the eastern Atlantic region as a precursor to extreme precipitation events \citep{toreti_precipitation_2016}. \cite{toreti_characterisation_2010} show that over the Western Mediterranean region, the negative geopotential height anomaly pattern associated with extreme precipitation is associated with an alignment of the subtropical jet with the African coastline and anomalous southwesterly surface to mid-tropospheric flow leading to large-scale ascending motions and instability over the Western Mediterranean region.

Weather regimes over both the North Atlantic and European, as well as the Mediterranean region, have been reported to modulate the occurrence of extreme precipitation over Morocco \citep{driouech_weather_2010, gadouali_link_2020, giuntoli_seasonal_2022, mastrantonas_extreme_2020, pasquier_modulation_2019}. For example, \cite{mastrantonas_forecasting_2022} demonstrate that Mediterranean weather regimes determine a significant increase in the probability of above 95th percentile precipitation. Using this information in a simple hybrid forecasting approach, they are able to slightly improve medium-range forecast skill over Morocco. \cite{gadouali_link_2020} on the other hand identify seven wintertime weather regimes over the North Atlantic and show their association with precipitation over Morocco, along with their modulation by the Madden-Julian Oscillation. Recent findings by \cite{chaqdid_extreme_2023}, investigating geopotential height, vertically integrated water vapour flux and wind speed anomalies associated with precipitation extremes over Morocco, however, indicate that these dynamical precursors might not be optimally resolved in weather regimes over either of the two regions, highlighting the scope for a more targeted approach in this region.

\subsection{Research gap: identifying targeted weather regimes}

Weather regimes are commonly identified using a combination of dimensionality reduction and clustering methods. While the dimensionality reduction step projects the high-dimensional data into a lower-dimensional subspace, the clustering subsequently identifies and assigns discrete regimes within this reduced space \citep{hannachi_low-frequency_2017}. 

\begin{figure}[!]%
\centering
\includegraphics[width=0.95\textwidth]{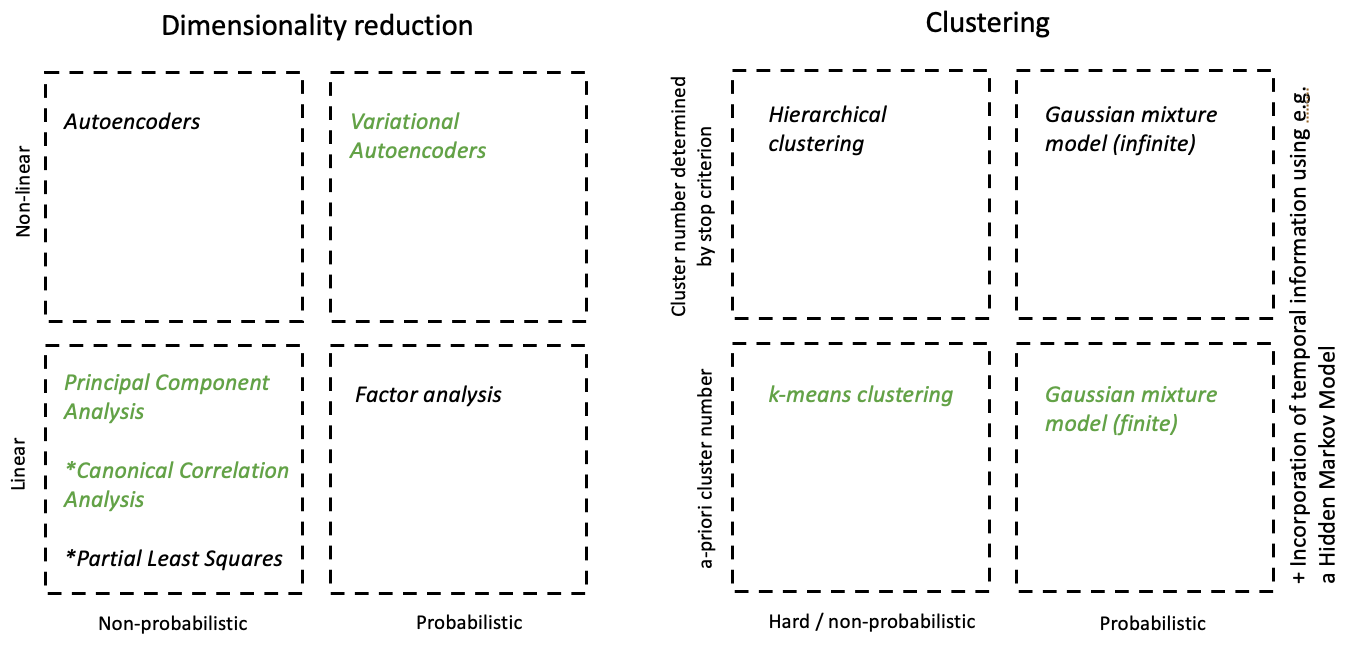}
{\caption{Illustration of selected methodological choices for dimensionality reduction and clustering based on \cite{hannachi_low-frequency_2017} and \cite{murphy_probabilistic_2022}. The methods highlighted in green are applied in this paper and described in more detail in section \ref{methods}. Methods highlighted with a star refer to joint dimensionality reduction methods between two high-dimensional spaces.}
\label{methods_overview}}
\end{figure}

Following \citet{michelangeli_weather_1995}, Principal Component Analysis (PCA, often referred to as Empirical Orthogonal Function or EOF analysis in atmospheric sciences \citep{hannachi_empirical_2007}) and k-means have established themselves as common choices for dimensionality reduction and clustering and have been applied in the relevant studies investigating weather regimes over the Western Mediterranean region \citep{gadouali_link_2020, mastrantonas_extreme_2020}. The advantage of this combination of methods is that they are easy to compute, understand, and interpret. However, they are not inherently more physically meaningful than other statistical dimensionality reduction and clustering methods. Alternative methods have been proposed in the literature on weather regimes, addressing, for example, the non-probabilistic nature of both methods which can lead to a loss of information on transitional states \citep{falkena_bayesian_2023}. Figure \ref{methods_overview} provides a summary of possible choices of dimensionality reduction and clustering methods for the identification of weather regimes. For a detailed discussion of existing methods for identifying weather regimes, we refer to \cite{hannachi_low-frequency_2017} and \cite{franzke_atmospheric_2017}.

The relationship between weather regimes and surface-level variables provides a key motivation for their investigation. However, most methods for identifying weather regimes are not specifically designed to capture variations of the impact variable in question, such as extreme precipitation over Morocco in this application. Therefore, available methods do not necessarily resolve the dynamical processes that modulate the relevant surface-level impact. Due to this limitation, studies investigating dynamical precursors of extremes using weather regimes might fail to capture relevant physical processes as well as potential predictors at subseasonal-to-seasonal lead times.

Recognizing the value of atmospheric patterns that are more informative of a local-scale variable, existing methods for identifying targeted atmospheric patterns are based on either pre-filtering data to extreme impact days \citep{dorrington_domino_2024, dorrington_precursors_2024, rouges_european_2023}, clustering the impact variable directly \citep{bloomfield_characterizing_2020, ullmann_euro-atlantic_2014}, or increasing the number of clusters to maximize the informativeness of regimes regarding the impact variable \citep{gadouali_link_2020, mastrantonas_extreme_2020}. However, these approaches compromise either regime persistence and robustness, and thereby their extended-range predictability, or the completeness of the representation of atmospheric dynamics. On the other hand, linear statistical methods for identifying related subspaces of two high-dimensional datasets such as canonical correlation analysis (CCA) have been applied, amongst others, by \cite{vrac_weather_2010} in combination with k-means clustering to identify weather regimes targeted to rainfall over France. However, CCA, for example, identifies linear transformations such that the two reduced spaces are maximally correlated \citep{murphy_probabilistic_2022}, thereby projecting the data into partial subspaces and compromising the ability of the regimes to represent the full atmospheric phase space.

Both the optimal number of clusters \citep{dorrington_jet_2020, falkena_revisiting_2020, franzke_atmospheric_2017}, as well as the physical and statistical interpretation of the weather regimes \citep{hochman_atlantic-european_2021, stephenson_existence_2004} have been subject to discussion, in particular, whether a multi-modality of the underlying probability density function is assumed. In this paper, weather regimes are here interpreted as statistical representations of the underlying physical processes that should be statistically robust and relevant to the intended use case, without making any stronger assumptions about the multi-modality of the underlying probability density function.

\subsection{Contribution}

To address the research gaps outlined above, this paper presents a novel method for identifying probabilistic weather regimes targeted to a local-scale scalar impact variable based on a modified variational autoencoder architecture. The proposed method, called RMM-VAE (Regression Mixture Model Variational Autoencoder), combines targeted dimensionality reduction with probabilistic clustering. This is achieved by integrating a regression into the dimensionality reduction step of the variational autoencoder and regularizing the reduced space using a Gaussian mixture model. The method thereby aims to capture the dynamical processes that modulate the target variable while maintaining the physical robustness and persistence of the identified regimes.

The regimes identified by the RMM-VAE method are probabilistic as each datapoint is assigned probabilities of belonging to the different clusters, and the clusters themselves are fit as multi-dimensional Gaussian distributions. One advantage of probabilistic clusters compared to so-called hard clusters identified for example by k-means is that information on transitional states can be captured, leading to a more complete picture of reduced atmospheric dynamics \citep{falkena_bayesian_2023}.

Variational autoencoders (VAEs) are a deep generative machine learning method introduced by \cite{kingma_auto-encoding_2013} and described in more detail in section \ref{methods}, that have shown promise in identifying non-targeted weather regimes \citep{baldo_probabilistic_2022}. The advantages of using a variational autoencoder architecture for identifying \textit{targeted} weather regimes lie in their ability to generalize the linear dimensionality reduction conducted in PCA to nonlinear transformations while offering the possibility of fitting an extendable probabilistic model in the dimensionality-reduced space. The approach presented here builds on previous machine learning architectures reported by \cite{zhao_variational_2019-1}, abbreviated R-VAE (Regression - VAE) that target the dimensionality reduction of a VAE without combining it with clustering, and \cite{ye_mixtures_2020} who fit a mixture model into the reduced space of a VAE. 

The RMM-VAE method is applied to identify weather regimes over the Mediterranean region in extended winter (November to March) targeted to total precipitation over Morocco and is compared to two established linear approaches (PCA + k-means, and CCA + k-means), as well as the R-VAE method \citep{zhao_variational_2019-1} combined with k-means clustering (R-VAE + k-means) which is introduced in section \ref{methods}.

The performance of these four methods is analysed in terms of the predictive skill of the resulting regimes with respect to the target variable, as well as their persistence and separability. These evaluation metrics are chosen to assess both the ability of the regimes to capture the relevant dynamical processes modulating the target variable and their physical robustness. To enhance the understanding and interpretability of the novel methods, we further investigate the reduced space - also called the latent space - along with the ability of the different dimensionality-reduction methods to reconstruct the input space from the reduced representation.

The remainder of this paper is structured as follows. Section 2  describes the data used, Section 3 provides a detailed description of the different methods, including the RMM-VAE method, and Section 4 reports details of the implementation and parameter choices. After comparing the results of the different methods for identifying targeted weather regimes in Section 5, Section 6 discusses and concludes the findings of this paper.

\section{Data}\label{sec:data}

Atmospheric circulation patterns are investigated over the Mediterranean region (lat: 25°N - 50°N; lon: 20°W - 45°E - region shown in figure \ref{cluster_centers}) in extended winter (Nov - Mar) using ERA5 reanalysis data from 1940 to 2022 \citep{hersbach_era5_2020} of geopotential height at 500hPa (z500) re-gridded to a resolution of 2.5°x2.5°. The data is standardized by subtracting the climatological daily mean and dividing the result by the standard deviation at each grid point. While this breaks the geostrophic relationship between geopotential height and the horizontal wind field, the dimensionality-reduction and clustering methods do not make use of this relationship, and the choice was therefore deemed acceptable for this application.

\begin{figure}[!]%
\centering
\includegraphics[width=0.9\textwidth]{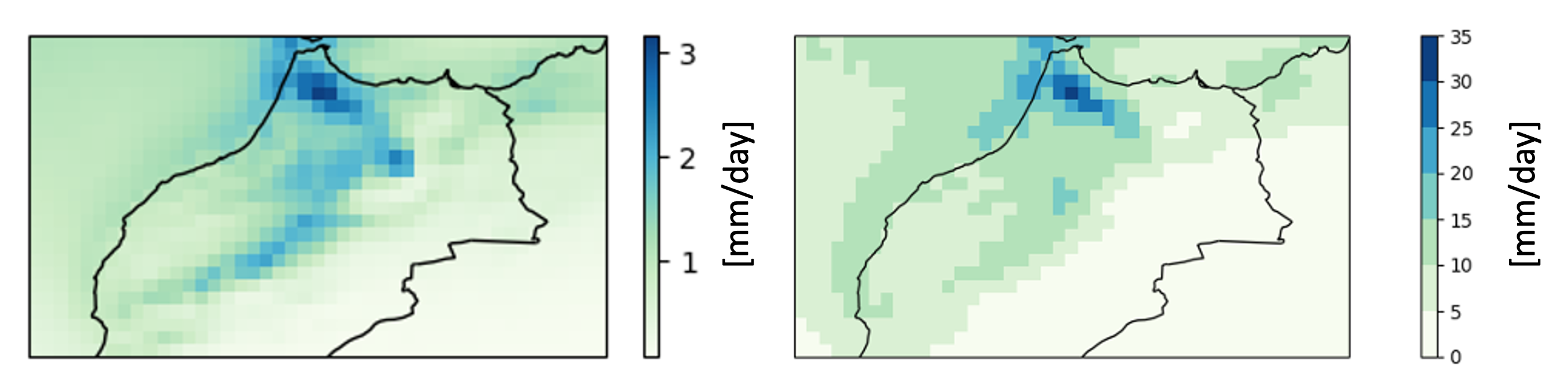}
{\caption{Extended winter precipitation (November - March) over the selected region over Morocco based on ERA5 reanalysis data from 1940-2022. Left-side: daily mean precipitation. Right-side: 95th percentile of daily precipitation}
\label{morocco}}
\end{figure}

For the initial application of the new method, ERA5 reanalysis data of daily total precipitation in a region over Morocco (lat: 30°N - 36°N; lon: 11°W - 0°E, resolution: 0.25°x0.25°) is used as a target variable over the same period as a proxy for observations. Figure \ref{morocco} shows the annual mean and 95th-percentile precipitation over the selected region. Precipitation data was normalized by applying a Box-Cox transformation, at each grid cell for CCA, and over the entire region for the two VAE methods \citep{box_analysis_1964}. A three-day running average of daily total precipitation and a five-day running average of z500 were taken to reflect the duration of extreme precipitation events and associated weather systems during the observational period \citep{dayan_review_2015, loudyi_flood_2022}.

\section{Methods}\label{methods}

This study compares four different methods for identifying weather regimes. All methods combine dimensionality reduction of the geopotential height space $\vect{x}$ into a latent space $\vect{z}$ with subsequent clustering. Except for the first method, PCA + k-means, all methods explicitly make use of the target variable $\vect{t}$, precipitation over Morocco, in the identification of weather regimes. 

\subsection{Principal component analysis and k-means clustering (PCA + k-means)}\label{pca_description}

The PCA + k-means method combines linear dimensionality reduction using principal component analysis (PCA) with subsequent clustering using k-means to identify weather regimes \citep{michelangeli_weather_1995}. PCA provides a linear transformation of the data into a subspace $\vect{z}$ spanned by the orthogonal eigenvectors of the covariance matrix \boldmath$C_{xx}$ of the dataset $\vect{x}$ \citep{jolliffe_principal_2016}. k-means clustering is applied to iteratively partition the reduced space $\vect{z}$ into \unboldmath$k$ sets with the objective of minimizing the within-cluster squared distance from the cluster centre \citep{murphy_probabilistic_2022}. This method is implemented due to its prevalence in the current literature, including existing studies on Mediterranean weather regimes and precipitation over Morocco \citep{mastrantonas_extreme_2020}. 

\subsection{Canonical correlation analysis and k-means clustering (CCA + k-means)}

The CCA + k-means method also combines a linear dimensionality reduction method with k-means clustering. Canonical correlation analysis (CCA) is a dimensionality reduction method that identifies linear transformations of two high-dimensional spaces, \boldmath$x$ and \boldmath$t$, into respective subspaces such that the correlation between the projections of the variables onto their new basis vectors is maximized \citep{johnson_applied_2013}. The method is symmetric, meaning the spaces \boldmath$x$ and \boldmath$t$ are treated in the same manner. In contrast to the machine learning methods presented in this paper, CCA takes the full precipitation field as input, rather than the aggregate scalar total precipitation over the selected region. CCA is applied to the input geopotential height space \boldmath$x$ and target precipitation space \boldmath$t$. Subsequently, weather regimes are identified by applying the k-means clustering algorithm to this dimensionality-reduced geopotential height space. This combination of methods is implemented to identify targeted clusters based on an established linear dimensionality reduction method.

\subsection{Regression variational autoencoder (R-VAE + k-means)}\label{rvae_method}

The R-VAE + k-means method is a targeted approach to the identification of weather regimes. The method is an extension of a variational autoencoder (VAE) architecture that combines dimensionality reduction with a prediction task introduced by \cite{zhao_variational_2019-1} in a neuroscience application. 

In a first application to climate data, this approach is amended to target the dimensionality reduction of the geopotential height input space $\vect{x}$ into the latent space $\vect{z}$ to the impact variable $\vect{t}$, i.e. total precipitation over Morocco. Weather regimes are identified by subsequentially clustering the reduced space $\vect{z}$ using the k-means clustering algorithm described in section \ref{pca_description}.

Sub-section \ref{vae_description} introduces variational autoencoders and variational inference. Any foundational technical statistical and machine learning terminology not explained directly in the text is highlighted in italics and introduced in more detail in a glossary in Appendix \ref{glossary}. Sub-section \ref{rvae_description} then provides a detailed description of the R-VAE method and its application to identify targeted weather regimes.

\begin{figure}[!]%
\centering
\includegraphics[width=0.9\textwidth]{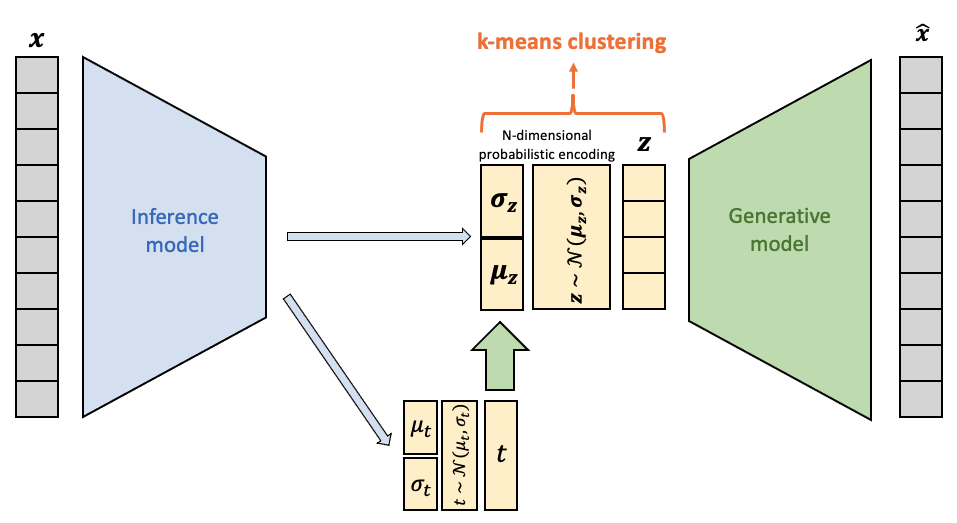}
{\caption{Schematic diagram of the R-VAE + k-means method, based on the architecture developed by \cite{zhao_variational_2019-1}. The input data $\vect{x}$ is passed through the encoder network, shown here in blue, which outputs both an estimate of the latent space $\vect{z}$ as well as a prediction of the scalar target variable $\vect{t}$. The target variable is then used to predict back into the latent space, thereby targeting the dimensionality reduction. The reduced space is subsequently clustered using k-means}
\label{rvae_graphic}}
\end{figure}

\subsubsection{Variational Autoencoders}\label{vae_description}

Autoencoders can be interpreted as a non-linear extension of PCA implemented through an encoder and decoder neural network. In PCA, the encoder would correspond to the linear transformation of the high-dimensional input data into the dimensionality-reduced space of principal components, while the decoder corresponds to the inverse transformation which reconstructs the input space using a selected number of principal components. Although autoencoders are more efficient at encoding the input data compared to PCA, the identified \textit{latent} space is not necessarily continuous which is an obstacle to subsequent clustering \citep{murphy_probabilistic_2022}. 

Variational autoencoders (VAEs) are a generative machine learning architecture introduced by \citet{kingma_auto-encoding_2013}. They extend the encoder-decoder architecture of autoencoders by fitting a probabilistic model of the data into the reduced space using Bayesian variational inference. By estimating the underlying probability distribution of the data in the latent space explicitly, the architecture allows for the generation of new samples from the encoded data, hence the term generative model. VAEs thereby provide a probabilistic and non-linear dimensionality reduction method and alternative to PCA.

The probabilistic graphical model underlying the variational autoencoder (VAE) architecture is shown in Figure \ref{graphical_models}a. The model aims to identify a continuous latent space \boldmath$z$ that provides a dimensionality-reduced representation of the high-dimensional input space \boldmath$x$. This statistical model with parameters $\theta$ can be fit using Bayesian inference based on Bayes theorem. However, this requires computing the \textit{posterior probability} $p_{\theta}(\vect{z}\mid \vect{x})$ which is in general computationally intractable. Therefore, the loss function of a variational autoencoder is derived using Bayesian variational inference \citep{murphy_probabilistic_2023}. Variational inference introduces a function $q_\phi$ from a selected distributional family with parameters $\phi$ to approximate the intractable posterior by minimizing the \textit{Kullback-Leibler (KL) divergence} between the true and approximated posterior. Those terms of the KL-divergence that depend on the parameters of the model represent a lower bound to the likelihood of the data and are termed the Evidence Lower Bound $\mathcal{L}(\theta, \phi \mid \vect{x})$. This Evidence Lower Bound can then be minimized to provide the best variational estimate of the model, using the Stochastic Gradient Variational Bayes (SGVB) estimator in the case of variational autoencoders. \citet{kingma_auto-encoding_2013} further introduce the reparameterization trick to generate samples from the probabilistic encoder $q_\phi (\vect{z} \mid \vect{x})$ while still being able to \textit{backpropagate} information through the network.
\begin{align}\label{vae_loss}
\mathcal{L}(\vect{x}) = - \mathbb{E}_{q_\phi(\vect{z} \mid \vect{x})}\left[ \log p_\theta(\vect{x} \mid \vect{z}) \right] + D_{KL} \left(q_\phi (z \mid \vect{x}) \mid p_\theta (z) \right).
\end{align}
Equation \ref{vae_loss} shows the resulting loss function that is minimized to fit a standard variational autoencoder. The first term represents the reconstruction loss of passing data points through the encoder and reconstructing it from its reduced representation. The second term represents the \textit{regularization} loss which penalizes the divergence of the fitted probability distribution in the latent space \boldmath$z$ from the \textit{prior probability distribution} $p_\theta (z)$, which is often assumed to be a multivariate Gaussian with mean $\mu_z$ and standard deviation $\sigma_z$.

\subsubsection{The R-VAE method}\label{rvae_description}

\begin{figure}[!]%
\centering
\includegraphics[width=0.98\textwidth]{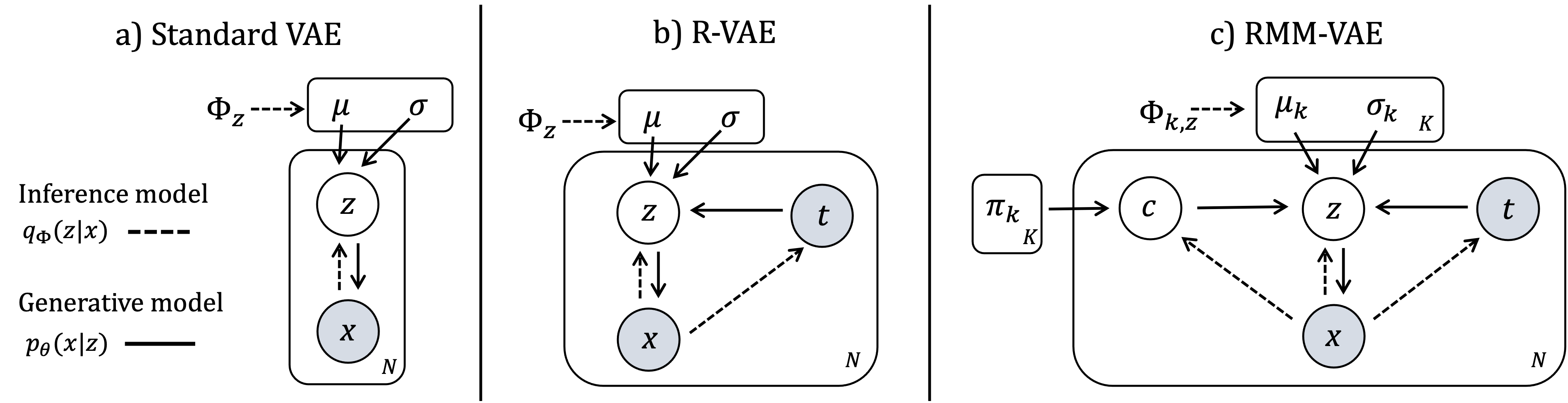}
{\caption{Different variational autoencoder models represented as probabilistic graphical models using plate notation. The inference model corresponds to the encoder and the generative model to the decoder of the architecture. a) A standard VAE with input variable $\vect{x}$, latent variable $\vect{z}$ and prior \unboldmath$\Phi_z$ on the parameters $\mu$ and $\sigma$ of the multivariate Gaussian distribution of $\vect{z}$. b) The R-VAE method with an additional target variable $t$, and c) the RMM-VAE method with probabilistic cluster assignment $\vect{c}$ regularized by the prior $\pi_k$. In all panels, dashed lines indicate the inference model and solid lines the generative model}
\label{graphical_models}}
\end{figure}

\cite{zhao_variational_2019-1} demonstrate that the standard VAE architecture can be extended to not only dimensionality-reduce the input space $\vect{x}$ but also predict a scalar target $\vect{t}$ variable that is subsequently used to \textit{regularize} the latent space \boldmath$z$. The method, termed R-VAE in this application, is illustrated in Figures \ref{rvae_graphic} and \ref{graphical_models}b. Here, the inference model, shown in blue in Figure \ref{rvae_graphic} and dashed lines in Figure \ref{graphical_models}b, not only estimates a dimensionality reduced space $\vect{z}$ from the input space $\vect{x}$ but also predicts the mean and variance of a target variable $\vect{t}$. In the generative model, shown in green in Figure \ref{rvae_graphic} and solid lines in Figure \ref{graphical_models}b the target variable is then used to predict back into the latent space, thereby providing an additional regularization of the reduced space. The impact of this additional regularization when applying this model to the dimensionality reduction of geopotential height using precipitation over Morocco as a target variable is evaluated in section \ref{latent}.

The loss function of the model can be derived using the representation as a probabilistic graphical model shown in Figure \ref{graphical_models}b to provide the following factorization of the joint probability distributions:
\begin{align}
p_\theta(\vect{x}, \vect{z}, t) = p_\theta(\vect{x} \mid \vect{z}) p_\theta(\vect{z} \mid t) p_\theta(t) \text{, and } q_\phi (\vect{z}, t \mid \vect{x}) = q_\phi (\vect{z} \mid \vect{x}) q_\phi (t \mid \vect{x}).
\label{rvae_jointp}
\end{align}
The term $p_\theta(\vect{x} \mid \vect{z})$ corresponds to the input space $\vect{x}$ reconstructed from the latent space $\vect{z}$ through the decoder network, and the term $p_\theta(\vect{z} \mid \vect{t})$ to the latent space estimated from the predicted target variable $\vect{t}$, while the prior distribution $p_\theta(t)$ here simply corresponds to the ground truth data of the target variable. The term $q_\phi (t \mid \vect{x})$ corresponds to the regression of the target variable from the input space $\vect{x}$, and the term $q_\phi (\vect{z} \mid \vect{x})$ to the latent space estimated from the input space using the encoding network.

The loss function of this modified VAE can then be derived as the KL divergence of the two probability distributions. For the full derivation see \cite{zhao_variational_2019-1} and appendix \ref{appendixloss}.
\begin{align}\label{rvae_loss}
\mathcal{L}(\vect{x}) &= -D_{KL} \left(q_\phi(\vect{z}, t \mid \vect{x}) \mid p_\theta(\vect{z}, \vect{x}, t) \right) \\
&= \mathbb{E}_{q_\phi(\vect{z} \mid \vect{x})}\left[ \log p_\theta(\vect{x} \mid \vect{z}) \right] - \mathbb{E}_{q_\phi(t \mid \vect{x})}\left[D_{KL}(q_\phi(\vect{z} \mid \vect{x}) \mid p_\theta(\vect{z} \mid t))  \right] \nonumber \\ &- D_{KL} \left(q_\phi (t \mid \vect{x}) \mid p_\theta (t) \right). \nonumber
\end{align}

As in the standard VAE described in the previous section, the first term represents the reconstruction loss of the dimensionality reduction. The third term represents the regression loss term, penalizing divergence of the predicted target variable $q_\phi (t\mid \vect{x})$ from the ground truth data $p(t)$. The second term uses the predicted target variable to regularize the dimensionality-reduced input space by penalizing the divergence between the two estimates of the latent space \boldmath$z$: one which is based on the dimensionality reduction of the original geopotential height data, $q_\phi (\vect{z} \mid \vect{x})$, and one which is predicted from the precipitation target variable, $p_\theta (\vect{z} \mid t)$.

In this application, the two components of the inference model, $q_\phi (\vect{z} \mid \vect{x})$ and $q_\phi (t \mid \vect{x})$, are parametrized as N-dimensional Gaussian distributions and estimated using non-linear functions with parameters $\phi$. Similarly, the probabilistic decoder is parametrized as a Gaussian and modelled as a nonlinear function with parameters $\theta$. Under the assumption that the decoder captures the nonlinearity of the generative model, a linear model for $p_\theta (\vect{z} \mid t) \sim \mathcal{N}(\vect{a}t, \mathbb{I})$ is implemented, where $\vect{a}$ is a vector of unit norm. This constrains the number of parameters the model has to fit overall.

\subsection{Regression - Mixture Model Variational Autoencoder (RMM-VAE)}\label{rmmvae_section}

In the novel RMM-VAE method, we extend the R-VAE architecture to directly fit a Gaussian mixture model into the reduced space of the variational autoencoder, instead of fitting a single multi-dimensional Gaussian which is subsequently clustered using k-means as in the R-VAE + k-means approach. The RMM-VAE method thereby integrates probabilistic clustering and targeted dimensionality reduction in a single coherent statistical model. A conceptual advantage of this is the ability of the model to represent the different aims of targeted clustering - identifying physically robust as well as predictive clusters - as terms in a coherently derived loss function, hence allowing for their statistical interpretation and an investigation of their trade-offs.

\begin{figure}[!]%
\centering
\includegraphics[width=0.9\textwidth]{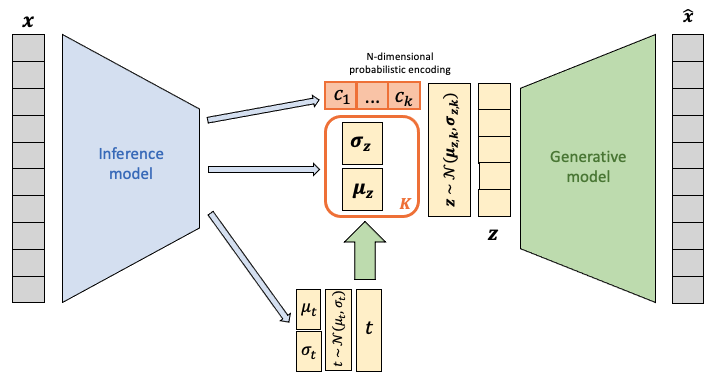}
{\caption{Schematic diagram of the proposed RMM-VAE approach. In contrast to the R-VAE method, the encoder network, shown again in blue, outputs not only an estimate of the latent space $\vect{z}$ and a prediction of the scalar target variable $t$, but also a probabilistic cluster assignment of the data point $c_k$. The method thereby combines a regression VAE (R-VAE) with probabilistic clustering using mixture models (MM) in a coherent statistical model}
\label{rmmvae_graphic}}
\end{figure}

To derive the RMM-VAE method, the single multi-dimensional Gaussian chosen to regularize the latent space in the R-VAE method is replaced by $k$ multi-dimensional Gaussians with mean $\vect{\mu_k}$ and standard deviation $\vect{\sigma_k}$. In addition, the probabilities $c_{ik}$ of the datapoint $x_i$ belonging to cluster $k$ are estimated. This builds on architectures combining variational autoencoders with mixture models presented, for example, by \cite{ye_mixtures_2020}, \cite{zhao_variational_2019} and \cite{jiang_variational_2017}. Gaussian mixture models themselves are an established probabilistic clustering method \citep{murphy_probabilistic_2022} that have been used to identify probabilistic weather regimes \citep{franzke_atmospheric_2017, baldo_probabilistic_2022}. Figure \ref{rmmvae_graphic} and Figure \ref{graphical_models}c illustrate the method. The inference model, shown again in blue in Figure \ref{rmmvae_graphic} and dashed lines in Figure \ref{graphical_models}c, estimates not only the dimensionality reduced space $\vect{z}$ and target variable $\vect{z}$ but also the cluster probabilities $c_{k}$.

The conditional independence assumptions embedded in the corresponding graphical model shown in Figure \ref{graphical_models}c are used to re-write the joint probability distribution of the model and derive the loss function of the model. For the full derivation of the loss function see appendix \ref{appendixloss}. 
\begin{align}\label{rmmvae_loss}
\mathcal{L}(\vect{x}) &= -D_{KL}(q_\phi (\vect{z},\vect{c},t \mid \vect{x}) \mid  p_\theta (\vect{x}, \vect{z}, t, \vect{c})) \nonumber \\
& = \sum_k q_\phi (\vect{c^k} \mid \vect{x}) \bigg[\mathbb{E}_{q_\phi (\vect{z} \mid \vect{x})} [\log p_\theta(\vect{x} \mid \vect{z})] - \mathbb{E}_{q_\phi (t \mid \vect{x})} [D_{KL} (q_\phi (\vect{z} \mid \vect{x}) \mid p(\vect{z} \mid t))] \nonumber \\
&- D_{KL} (q_\phi(t \mid \vect{x}) \mid p(t)) 
 - D_{KL} (q_\phi (\vect{z} \mid \vect{x}) \mid p_\theta(\vect{z} \mid \vect{c^k})) \bigg] \nonumber \\
 & -D_{KL}(q_\phi (\vect{c^k} \mid \vect{x}) \mid p(\vect{c^k})).
\end{align}

The first three terms in brackets correspond to the terms of the R-VAE loss function for an individual mixture component: the reconstruction loss, the divergence of the estimated target variable $q_\phi (t \mid \vect{x})$ from the ground truth data, and the divergence between the latent spaces generated from the target variable $p(\vect{z} \mid t)$ and the latent space encoded from the input data $q_\phi (\vect{z} \mid \vect{x})$), all weighted by the cluster assignment $q_\phi (\vect{c^k} \mid \vect{x})$. The fourth term of the loss function minimizes the divergence between the cluster mean k and the latent space estimated from the input data, again weighted by the probability of cluster k occurring, $q_\phi (\vect{c^k} \mid \vect{x})$. The final term regularizes the cluster assignment $q_\phi (\vect{c^k} \mid \vect{x})$ to the previous cluster occurrence frequency $p(c^k)$. 

The components of the inference model $q_\phi (\vect{z} \mid \vect{x})$ and $q_\phi (t\mid \vect{x})$, and generative model $p_\theta (\vect{x}\mid \vect{z})$ and $p_\theta (\vect{z} \mid t)$, are parametrized as in the previous method. $p(c^k)$ is a categorical distribution populated by the occurrence frequency of the different clusters updated at each step. This occurrence frequency is used as prior for the probabilistic cluster assignment of an individual day $q_\phi (\vect{c^k} \mid \vect{x})$. Individual mixture components $p(\vect{z}|\vect{c^k})$ are modelled as Gaussians with mean $\mu$ and the identity covariance matrix. The latter choice is made to constrain the number of parameters and avoid model overfitting.

\section{Experiments}

The encoders and decoders of both VAE methods are implemented using three dense layers of decreasing dimensionality of 128, 64 and 32 respectively. A batch size of 128 and the ReLU activation function are chosen. For 100 epochs, the model is trained on iterative train-test splits using k-fold cross-validation, and subsequently fitted again using a random weights initialization on the entire dataset. For the implementation of the neural network architectures, the Python package \textsc{keras} \citep{chollet_keras_2015} was used. For a number of the evaluation metrics as well as the implementation of the two linear methods, the Python package \textsc{scikit-learn} \citep{pedregosa_scikit-learn_2011} was employed.

Table \ref{tab} provides an overview of the compared methods and relevant hyperparameters. A 10-dimensional latent space was implemented for all the methods, and cluster numbers between 4 and 10 were investigated based on the understanding that the correct number of clusters will depend on the use case and cannot be determined in a general sense in all regions \citep{franzke_atmospheric_2017}. The sensitivity of the results to both these choices was investigated and, in the case of the cluster number, the sensitivity of the clusters to sub-sampling of the input data for different choices of k is shown in Appendix \ref{appendixA}. Based on these results, k=5 was identified as a reasonable choice of cluster number to visualise in the results section where required. 

For both VAE methods, the inclusion of a hyperparameter $\beta$ based on \cite {higgins_beta-vae_2017} is investigated. The hyperparameter is multiplied with the respective first terms in equations \ref{rvae_loss} and \ref{rmmvae_loss} representing reconstruction loss, thereby changing the weight of the reconstruction objective in the loss function. Two values of $\beta$ are explored for each of the two methods, whereby v1 ($\beta=1$) corresponds to the original loss function without the inclusion of an additional hyperparameter, and v2 ($\beta<1$) decreases the importance of the reconstruction loss term in the loss function. Selected values for $\beta$ are shown in table \ref{tab}.
\begin{table}[t]
\begin{center}
\caption{Overview of methods for the identification of weather regimes and associated parameter choices.\label{tab}}
\begin{tabular}{|l|l|l|l|}
\hline
\textbf{Method} & \textbf{Version} & \textbf{$\beta$ }& \textbf{Target variable} \\
\hline
PCA + k-means & - & -  & None \\
\hline
CCA + k-means & - & -  &$\vect{t}$ - full daily precipitation field at 0.25° resolution \\ \hline
R-VAE + k-means & v1 & 1  &$t$ - spatially averaged daily precipitation (scalar) \\
 & v2 & 0.1  & " - " \\ \hline
RMM-VAE & v1 & 1  & " - " \\
 & v2 & 0.5  & " - "  \\ \hline
\end{tabular}
\end{center}
\end{table}

\section{Results}

Section \ref{clusters} presents the weather regimes identified over the Mediterranean region using the different methods, alongside the conditional changes in the probability of extreme precipitation over Morocco. Next, the predictive skill of the regimes with respect to the target variable (section \ref{predictability}), and their physical robustness (section \ref{robust}) are evaluated and contrasted. To gain further insight and interpretability of the methods, we investigate the respective dimensionality-reduced spaces in section \ref{latent} and the reconstructed input spaces in section \ref{reconstructed}.

\subsection{Cluster centers and characteristics}\label{clusters}

\begin{figure}[!]%
\centering
\includegraphics[width=0.95\textwidth]{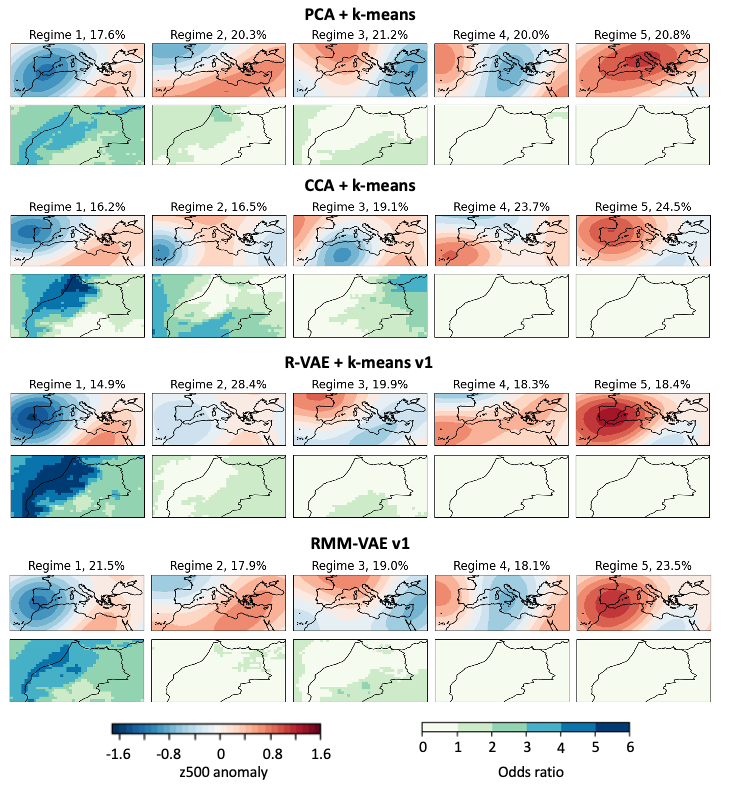}
{\caption{Identified weather regimes (top rows) and corresponding odds ratios of extreme precipitation (bottom rows) for the four different methods with the number of clusters specified as k=5. The regime frequencies are given in percent. The odds ratio of extreme precipitation corresponds to the ratio of the probability of the climatological 95th percentile of precipitation at the grid cell conditional on that weather regime, divided by the unconditional probability of 95th percentile of precipitation (i.e. 0.05). The weather regimes are ordered in decreasing order of total precipitation during the days assigned to this cluster by the respective method}
\label{cluster_centers}}
\end{figure}

Figure \ref{cluster_centers} shows the weather regimes, defined as the centres of the identified clusters, along with the associated change in the conditional probability of extreme precipitation over Morocco, defined as the exceedance of the grid cell-specific 95th percentile.

The weather regimes identified using the PCA + k-means method correspond to those found over the Mediterranean region in other publications using this method such as \cite{giuntoli_seasonal_2022} and \cite{mastrantonas_what_2022}. While the number of regimes investigated varies, both publications identify the meridional patterns observed in regimes 3 and 4, a high geopotential height anomaly (regime 5 - termed Mediterranean high in \cite{mastrantonas_what_2022}), and the western low anomalies seen in regimes 1 and 2, (termed Iberian and Biscay Low in \cite{mastrantonas_what_2022}). Regime 1, associated with a geopotential height low over the west of Europe, increases the probability of extreme precipitation by a factor of three to four while the other regimes show no or marginal increases in the probability of extreme precipitation. This is consistent with the results found by \cite{mastrantonas_extreme_2020}. Overall, the regimes identified by PCA + k-means have roughly similar frequencies of occurrence.

CCA + k-means, on the other hand, identifies multiple regimes associated with an increase in the conditional probability of extreme precipitation by a factor of three or more in different regions of Morocco (regimes 1-3). The spatial patterns of extreme precipitation appear to be modulated by the location of the geopotential height low around Morocco. In contrast, high geopotential height anomalies dominate over the western Mediterranean in the two regimes associated with a lower-than-average probability of extreme precipitation. The regimes associated with extreme precipitation (regimes 1-3) have a slightly lower frequency of occurrence than the other two (regimes 4-5). It can be observed that the anomalies that primarily define the CCA + k-means cluster centres are located in the Western Mediterranean region which will be further investigated in section \ref{reconstructed}.

In contrast to the different spatial patterns of extreme precipitation associated with different weather regimes identified using CCA + k-means, the R-VAE + k-means method identifies a single regime associated with a five to six times increase in the probability of extreme precipitation. Furthermore, we find that the cluster centre of regime 2, which occurs on almost 30\% of days, shows little mean z500 anomaly, meaning it is quite close to a climatologically average day in extended winter. This indicates that while the method is able to identify a regime which is highly informative regarding the occurrence of extreme precipitation, it might not be able to identify structure in the full atmospheric phase space. This will be further investigated in sections \ref{latent} and \ref{reconstructed}.

The regimes identified by the RMM-VAE method appear to strike a balance between the baseline method PCA + k-means, and the purely targeted method R-VAE + k-means. On the one hand, the method identifies a single regime associated with a higher probability increase of extreme precipitation compared to PCA + k-means, similar to R-VAE + k-means. On the other hand, the resulting cluster centres are visually more similar to the PCA + k-means cluster centres.

\begin{figure}[!]%
\centering
\includegraphics[width=0.9\textwidth]{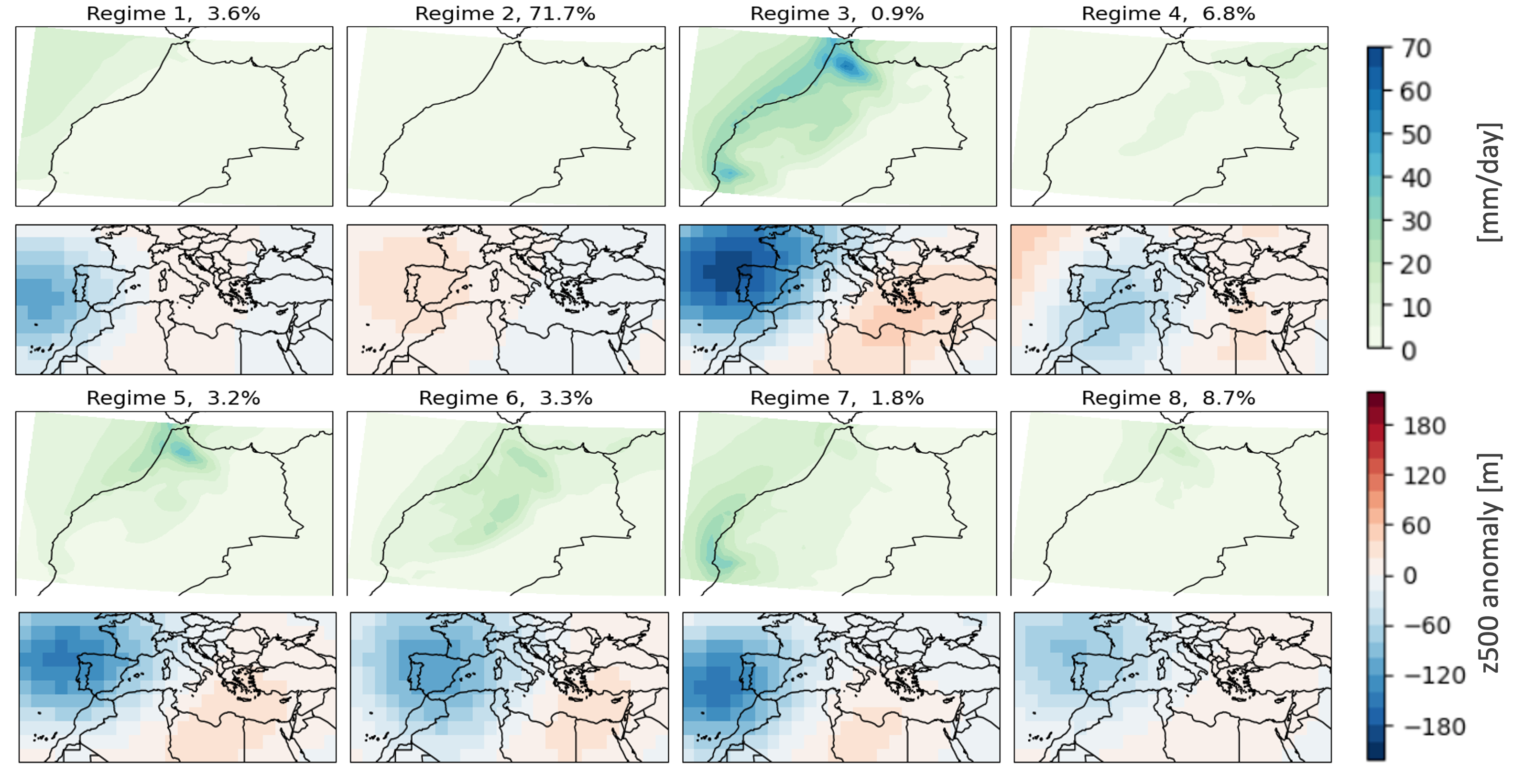}
{\caption{Precipitation clusters computed on precipitation reanalysis data without pre-processing using the k-means clustering algorithm for k=8 (top rows) and corresponding z500 anomalies (bottom rows)}
\label{precip_clusters}}
\end{figure}

To gain further insight into the dynamical precursors of extreme precipitation, we cluster the precipitation field directly using k-means clustering and investigate the corresponding average geopotential height anomalies, shown in figure \ref{precip_clusters}. As noted in section \ref{sec:intro}, clustering the impact variable directly compromises regime persistence and leads to an incomplete representation of the large-scale dynamics, and is therefore shown for illustration purposes only, but not compared as an alternative targeted clustering method. 

In agreement with the odds ratios associated with the weather regimes shown in Figure \ref{cluster_centers}, the location and intensity of the precipitation events appear to be modulated by the location and intensity of a geopotential height low off the coast of Spain, consistent with existing literature on dynamical drivers of extreme precipitation in the Western Mediterranean discussed in section \ref{morocco_section}. Comparing these patterns with the weather regimes shown in Figure \ref{cluster_centers}, we find that the two VAE methods cluster the different patterns of z500 anomalies identified in Figure \ref{precip_clusters} in one single weather regime, while CCA + k-means disaggregates some of the different spatial patterns of extreme precipitation into different weather regimes. This finding is consistent with the way the different methods incorporate the target variable: while CCA makes use of the entire precipitation field as input data and can hence extract more information about its spatial structure, the VAE methods only receive a single scalar target variable, total precipitation, as input, and are therefore not able to separate different spatial patterns.

\subsection{Evaluating the skill of weather regimes in predicting the target variable}\label{predictability}

To evaluate the ability of the weather regimes to capture the dynamical processes responsible for modulating extremes in precipitation over Morocco, we evaluate an empirical prediction of the target variable using the identified weather regime assignment. 

The prediction is calculated by multiplying the probability of the weather regimes assigned by the respective method with the conditional probability of the target variable given that weather regime. To assess the ability of the weather regimes to capture both the body and tail of the target variable distribution, both the skill of the weather regimes in predicting precipitation terciles and the exceedance of the 95th percentile are evaluated. The skill of this prediction is analysed using the Brier score (BS), a strictly proper scoring rule to measure the accuracy of a probabilistic prediction of mutually exclusive discrete outcomes widely used in forecast evaluation \citep{gneiting_strictly_2007}. Skill scores were also used by \cite{schiemann_how_2010} to quantify the surface impact of circulation types. The BS is defined as:
\begin{align}
BS = \frac{1}{N} \sum_{n=1}^{N} \sum_{j=1}^{m} (\delta_{i_n j} - p_j)^2 = - \frac{1}{N} \sum_{n=1}^{N} (1 - 2 p_{i_n} + \sum_{j=1}^{m} p_j^2),
\end{align}
where $m$ is the number of forecast categories and $N$ is the number of timesteps. $\delta_{i_n j}$ is the Kronecker delta which equals 1 if the observation $i$ at timestep $n$ corresponds to category $j$, and 0 otherwise, and $p_j$ the forecast probability of category $j$. The corresponding skill score (BSS) is calculated with respect to a reference forecast, chosen here to be the climatology over the entire period, and defined as
\begin{align}
BSS = 1- \frac{BS_{\text{forecast}}}{BS_{\text{climatology}}}.
\end{align}
A BSS of 1 indicates a perfect forecast while values close to zero indicate little, or no skill compared to the reference forecast. 

\begin{figure}[!]%
\centering
\includegraphics[width=0.9\textwidth]{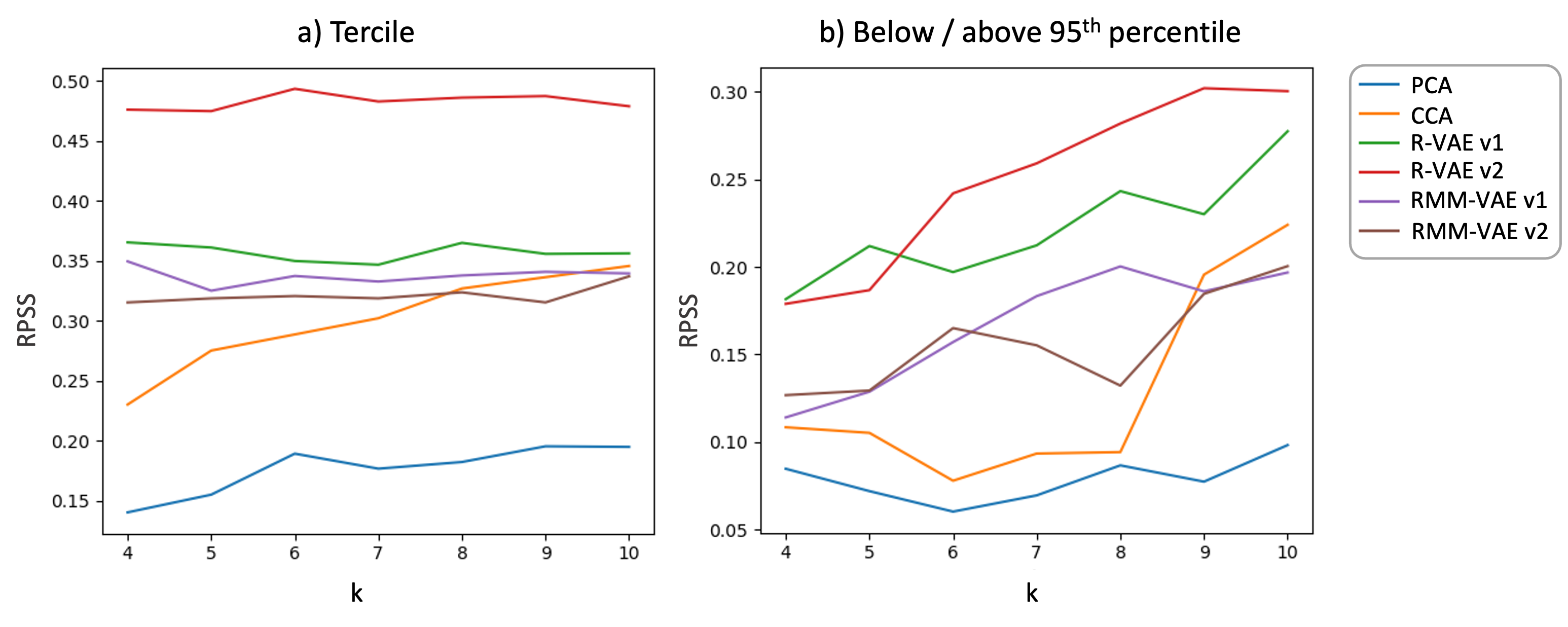}
{\caption{Brier skill score of an empirical prediction of total precipitation over Morocco using the weather regimes, shown for different numbers of weather regimes k. a) Skill score for the prediction of the tercile of the precipitation distribution, and b) Skill score for extreme precipitation, defined as a binary prediction above or below the 95th percentile. For probabilistic clustering, the skill score is computed using the most likely cluster at the given data point. The higher the BSS, the more predictive the weather regimes are of precipitation over Morocco}
\label{rpss}}
\end{figure}

The resulting BSS is calculated for precipitation terciles (Figure \ref{rpss}a), and for extreme precipitation (Figure \ref{rpss}b). The predictive skill is overall higher for terciles compared to the 95th percentile threshold, which is to be expected. In both cases, all targeted methods outperform PCA + k-means (blue line), highlighting the potential of improving the predictive skill of standard weather regime definitions. R-VAE + k-means (green and red lines) performs best, followed by RMM-VAE (purple and brown lines) up to k=8, after which it is slightly outperformed by CCA + k-means.

The better performance of R-VAE can be attributed to RMM-VAE having more objectives to achieve simultaneously as it aims to identify probabilistic clusters while also disentangling the latent space with respect to the target variable. For both VAE methods, increasing the importance of the prediction objective in the loss function (v2) further boosts the predictive skill, although not consistently across cluster numbers. 

Increasing the number of clusters does not improve the skill except for CCA + k-means. For the two VAE methods, this result is likely because the targeted dimensionality reduction already groups data points with similar precipitation amounts in the reduced space, as discussed in section \ref{latent}. Unlike the other methods, CCA takes the full precipitation field as input. The ability to extract spatial information about the precipitation field might enable the predictive skill to increase with cluster number.

\subsection{Evaluating the physical robustness of the weather regimes}\label{robust}

Existing literature on targeted weather regimes finds that patterns that are more informative of a local-scale target variable risk compromising the physical robustness of regimes which is associated with their persistence and subseasonal predictability. To evaluate the robustness of the targeted regimes, we assess the persistence and the separability of the identified circulation patterns as proxies for their physical robustness. 

The separability of the clusters is assessed using the silhouette score \citep{rousseeuw_silhouettes_1987}. The higher the silhouette score, the better the clusters are separated from one another, while a silhouette score close to zero indicates that the separation between different clusters is not statistically significant. The statistical robustness of the patterns to sub-sampling is also evaluated and shown in appendix \ref{appendixA}.

\begin{figure}[!]%
\centering
\includegraphics[width=0.9\textwidth]{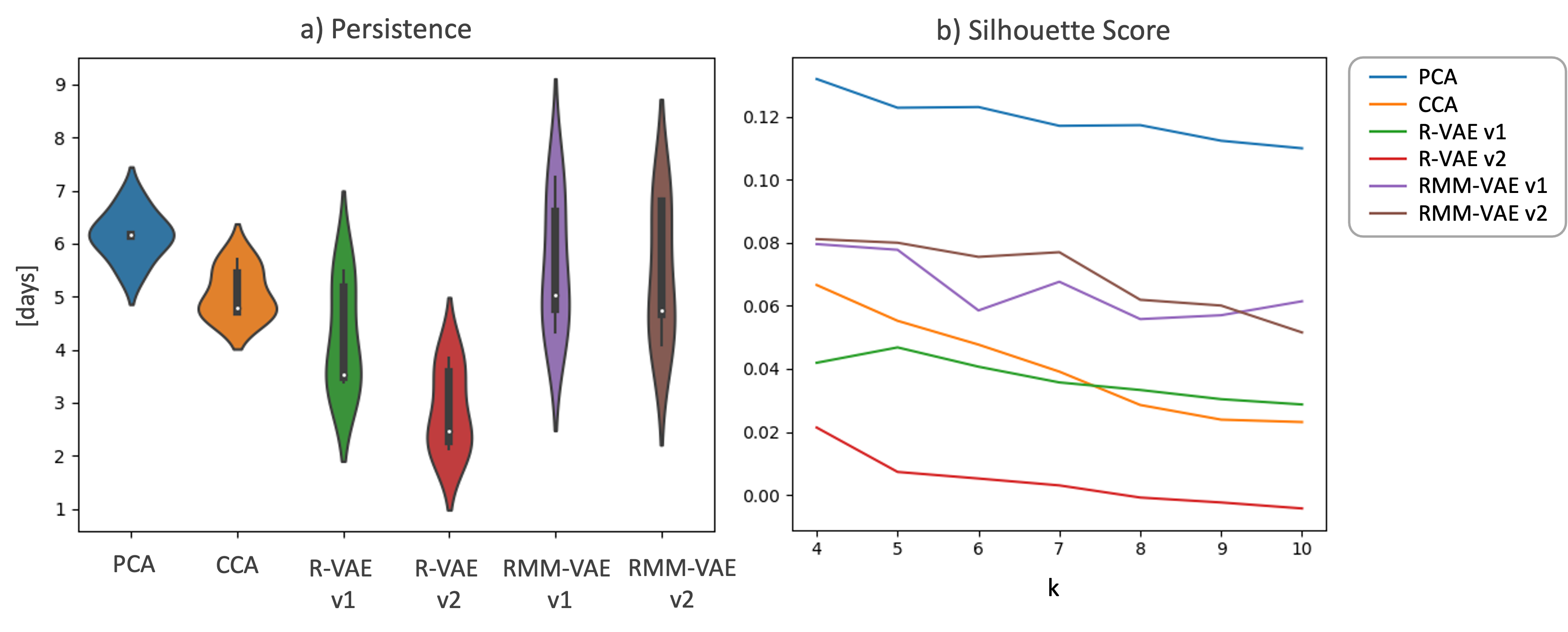}
{\caption{Regime persistence and separability. a) Distribution of mean persistence across $k=5$ regimes. Violin plots show the kernel density estimation of the distribution, the distribution median (white point) as well as the interquartile range (black box). The ranking of different methods in terms of persistence remains the same for different choices of $k$. b) Silhouette score for a range of cluster numbers k. The silhouette score defined as the mean silhouette coefficient (b-a)/max(a,b), where a is the average intra-cluster distance and b is the average inter-cluster distance i.e. the average distance between all clusters}
\label{persistence}}
\end{figure}

Figure \ref{persistence}a shows the distribution of mean cluster persistence across $k=5$ clusters. Mean persistence across clusters is highest for PCA, followed by RMM-VAE. However, the spread of the distribution of persistence across clusters is lower for PCA and CCA compared to the two VAE methods, in particular RMM-VAE. This result indicates that while all five PCA + k-means clusters have around the same average persistence, there are some clusters with a longer and some with a shorter average persistence identified by the VAE methods. These results are qualitatively similar for other choices of k (not shown). Sample time series of the cluster assignments in different methods are shown in appendix \ref{appendixB}. Similarly, all targeted clusters perform worse in terms of cluster separability (Figure \ref{persistence}b) compared to the baseline method PCA + k-means. RMM-VAE outperforms the other targeted methods while R-VAE performs the worst overall. 

Overall, these results indicate that RMM-VAE identifies more coherent and robust clusters compared to the other targeted methods, in particular the R-VAE + k-means method. This result indicates that there is a benefit in performing the probabilistic clustering in one coherent statistical model and predicting the target variable and dimensionality reducing the input space in a single step, as opposed to separating the two steps, as in the R-VAE + k-means method.

\subsection{Structure of the dimensionality-reduced spaces}\label{latent}

To enhance the interpretability of the methods, we investigate the dimensionality-reduced spaces identified by different methods along with the associated cluster assignment.

\begin{figure}[!]%
\centering
\includegraphics[width=0.99\textwidth]{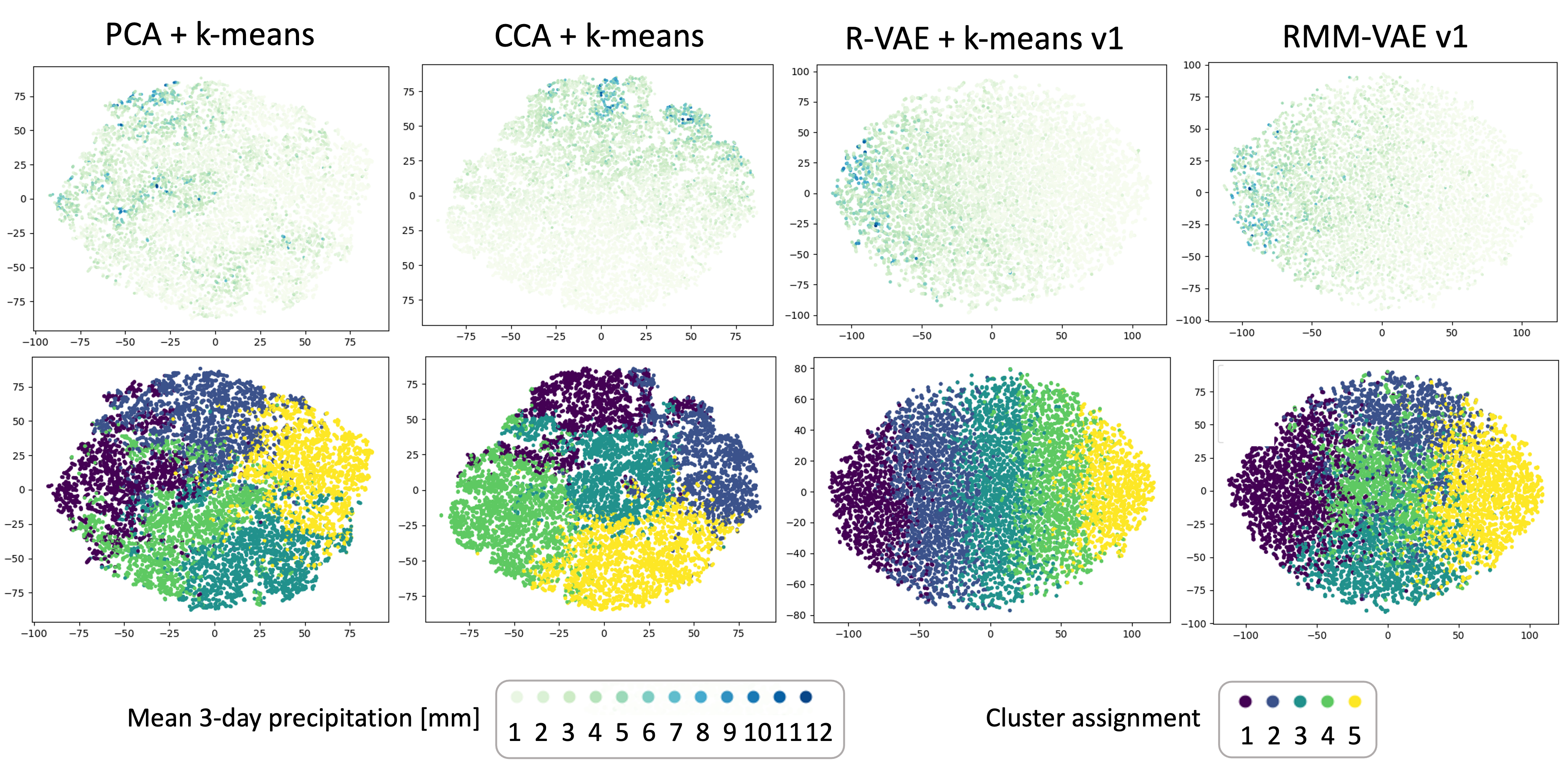}
{\caption{Visualisation of the 10-dimensional latent spaces in two dimensions using t-distributed stochastic nearest-neighbour embedding (t-SNE). Embedded data points are coloured according to the value of the target variable, total mean precipitation (top row), and according to the cluster they are subsequently assigned to (bottom row)}
\label{tsne}}
\end{figure}

All methods identify a 10-dimensional dimensionality-reduced space of geopotential height data of either principal components, canonical variates or multidimensional Gaussian distributions. To visualise this 10-dimensional representation in two dimensions, we use t-distributed stochastic nearest neighbour embedding (t-SNE) \citep{maaten_visualizing_2008}, a method commonly applied in the machine learning community for the visualisation of high-dimension datasets. The resulting visualisation, shown in Figure \ref{tsne}, provides an intuition for how the target variable is distributed in the dimensionality-reduced space (top row), and how different clustering methods capture this distribution (bottom row). The t-SNE method preserves nearest neighbours but projects the high-dimensional data onto dimensions that can no longer be interpreted in terms of the physical units of the original input space, therefore the axes in Figure \ref{tsne} do not have a unit associated with them. Different values of the perplexity parameter which determines the number of neighbours considered for each point were tested and a value of 10 was chosen as it shows representative results. When interpreting the results, it should be noted that the t-SNE method represents only one possible way of visualising the latent space.

We find that both targeted VAE methods, R-VAE and RMM-VAE, disentangle the dimension of the geopotential height dataset associated with variations in precipitation over Morocco, the scalar target variable $t$ (dark dots in Figure \ref{tsne}, top row). This aligns with the findings presented by \cite{zhao_variational_2019-1} demonstrating a similar type of disentanglement when applying the R-VAE method to the studied neuroscience application. Interpreting the disentanglement in the context of the dynamical drivers of extreme precipitation shown in Figures \ref{cluster_centers} and \ref{precip_clusters}, the disentangled dimension can be seen to represent the location and depth of the geopotential height anomaly over the Western Mediterranean shown to modulate precipitation over Morocco.

The PCA latent space, on the other hand, shows less organisation with respect to the target variable, as expected, since the target variable is not part of the dimensionality reduction. In contrast, the CCA latent space appears to have more structure compared to PCA, though not as aligned as the latent spaces identified by the R-VAE and RMM-VAE methods.

Colouring the data points according to their assigned cluster shown in the bottom row of figure \ref{tsne}, we find that the R-VAE and RMM-VAE methods achieve the improved predictive skill regarding the target variables shown in figure \ref{rpss} by first grouping data points associated with a similar precipitation impact in the dimensionality reduction step, and subsequently assigning them to one cluster. 

The R-VAE method, which carries out the targeted dimensionality reduction and clustering in two separate steps, identifies clusters in 'bands' along the dimension associated with the target variable. The RMM-VAE method on the other hand, which fits probabilistic clusters while simultaneously disentangling the dimension associated with the target variable, organises the clusters in a way that appears more consistent with the target of minimizing the distance between points in one cluster, which is also the case for PCA + k-means and CCA + k-means. This difference provides an interpretation and explanation as to why the RMM-VAE method is able to balance informativeness with respect to the target variable with cluster robustness better than R-VAE, although the latter provides a higher predictive skill.

\subsection{Ability to reconstruct the input spaces}\label{reconstructed}

To evaluate the performance of the dimensionality reduction performed by different methods, we compare the geopotential height data reconstructed from the dimensionality-reduced spaces to the original input data. In the case of PCA, for example, an input data point corresponding to a z500 anomaly pattern is compared to that same data point reconstructed using the first 10 principal components. In the case of the VAE methods, the input data point is compared to that same data point after passing it through the encoder and decoder of the model. 

\begin{figure}[!]%
\centering
\includegraphics[width=0.5\textwidth]{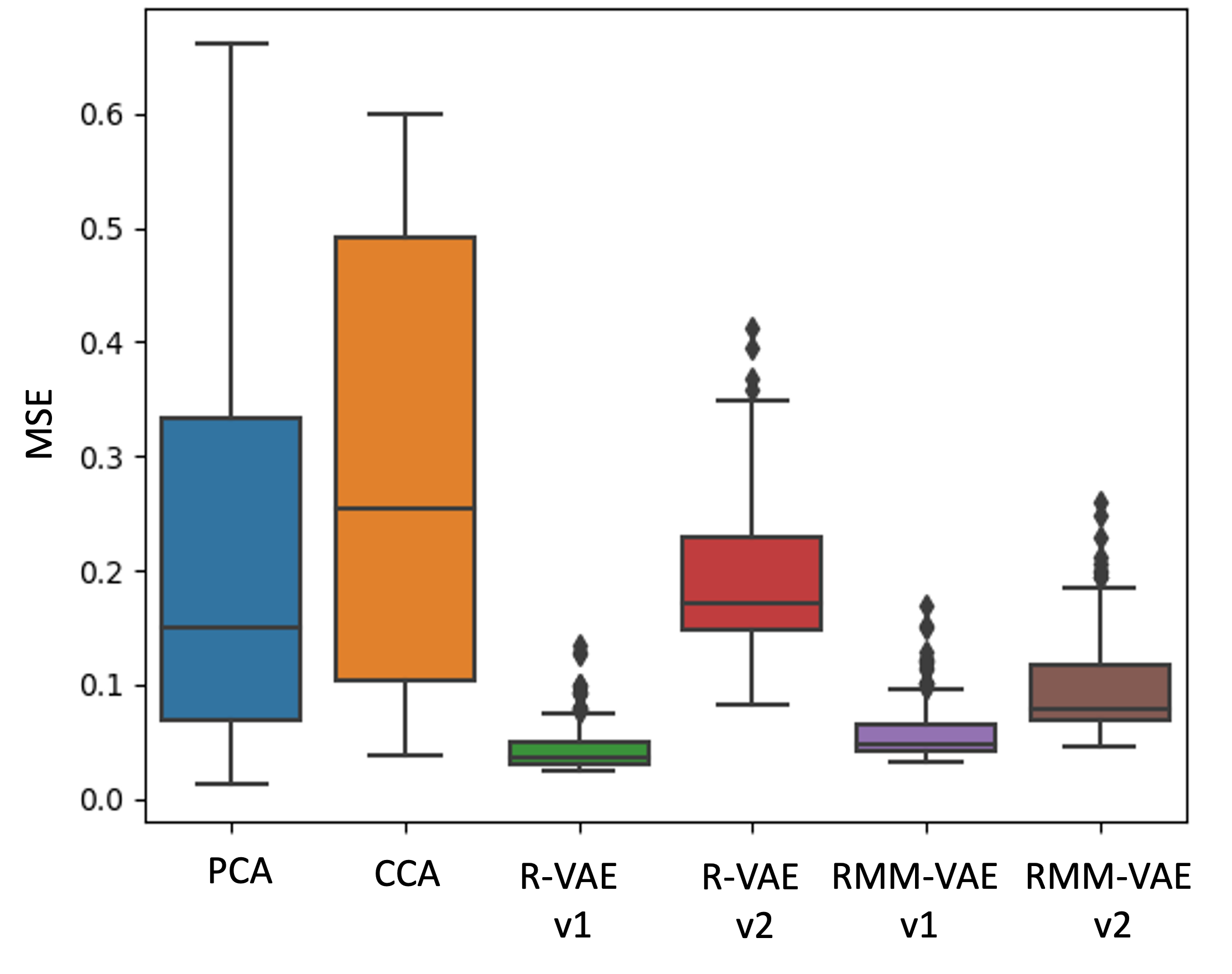}
{\caption{Distribution of the reconstruction loss, assessed using the mean squared error (MSE) between original input data and reconstructed data for all data points. The thin line in the boxes corresponds to the median of the distribution, while the boxes extend to 0.25 and 0.75 quartiles of the dataset. The whiskers extend to points that lie within 1.5 of the interquartile range of the lower and upper quartile. Observations outside this range are displayed as black points}
\label{rmse_boxplot}}
\end{figure}

The performance of the dimensionality reduction can then be assessed by computing the mean squared error (MSE) between input and reconstructed data.  The lower this value, the more information about the input space is captured by the dimensionality reduction method. Figure \ref{rmse_boxplot} shows the distribution of this error across all data points.

Both VAE methods, in particular the respective v1 methods (with $\beta$=1) have a lower and less widely distributed MSE compared to both PCA and CCA. This means that despite targeting their respective latent spaces to an impact variable, the VAE methods still outperform the two linear methods in terms of representing the atmospheric dynamics in a dimensionality-reduced space.

When increasing the importance of the prediction objectives in the VAE v2 models (with $\beta$<1), both R-VAE v2 and RMM-VAE v2 still outperform PCA and CCA but perform worse than their respective v1 counterparts. The RMM-VAE v2 ($\beta$=0.5) method performs slightly better than R-VAE v2 ($\beta$=0.1) which is consistent with the different values for $\beta$ chosen to ensure convergence of the model. This result shows that a trade-off between targeting the dimensionality reduction and reconstructing the full phase space exists: while both v2 methods perform better in the task of predicting the target variables, this comes at the cost of a loss of skill in the dimensionality reduction.

\begin{figure}[!]%
\centering
\includegraphics[width=0.9\textwidth]{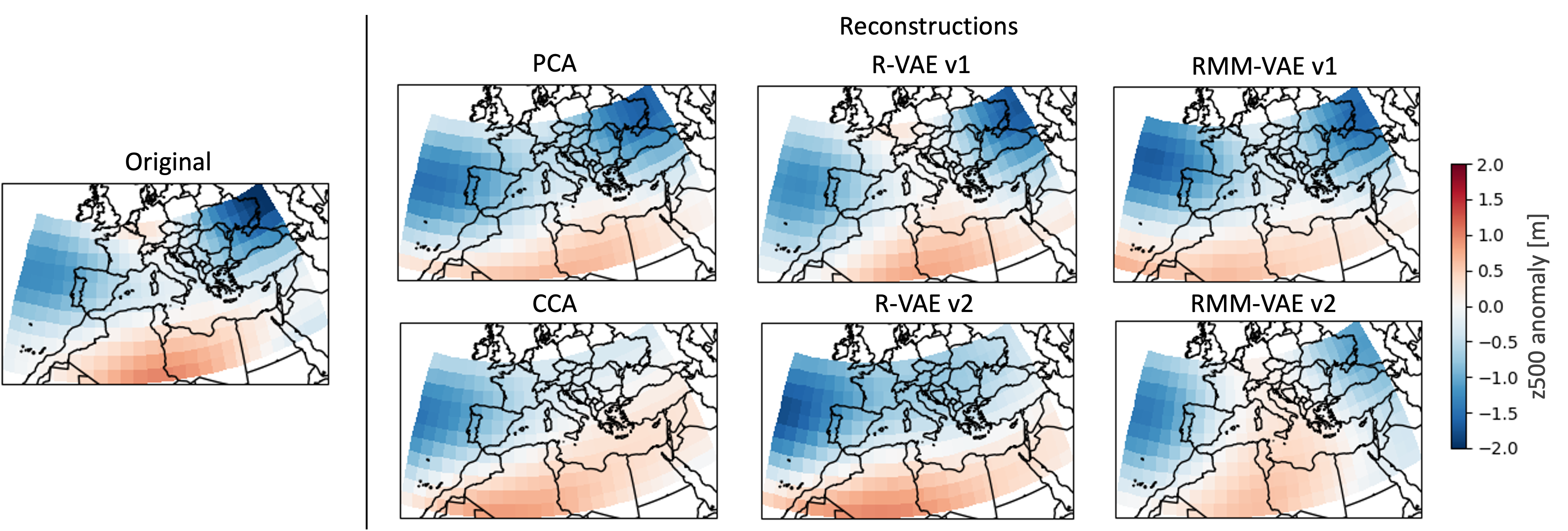}
{\caption{Gridded and normalized z500 anomalies, as detailed in Section 3, on an example day 1940-01-04, showing the original data on the left and the reconstructions using different methods on the right.}
\label{rmse_spatial}}
\end{figure}

Investigating the reconstruction of an individual data point, shown in Figure \ref{rmse_spatial}, we find that both v2 methods and CCA focus the dimensionality reduction on the Western Mediterranean region surrounding Morocco. This explains the worse performance of the two v2 methods compared to their respective v1 versions, as well as the worse performance of CCA compared to PCA in Figure \ref{rmse_boxplot}. The result highlights that some methods for identifying targeted weather regimes such as CCA come at the cost of only representing the dynamics of a partial subspace. The two VAE v1 methods on the other hand, appear to be able to balance this trade-off.

\section{Discussion and Conclusion}

In this paper, we present a novel machine learning method, RMM-VAE (Regression Mixture Model Variational Autoencoder), for identifying weather regimes targeted to a scalar impact variable. The method combines the different objectives of targeting weather regimes - predicting the target variable and identifying robust dynamical patterns - in a coherent probabilistic model for the first time. The model integrates non-linear and targeted dimensionality reduction with probabilistic clustering using Gaussian mixture models in a modified variational autoencoder architecture described in section \ref{rmmvae_section}.

The RMM-VAE method is applied to identify weather regimes over the Mediterranean region targeted to precipitation over Morocco. Results are compared to three alternative approaches: principal component analysis (PCA) combined with k-means clustering as the currently established standard practice for identifying weather regimes \citep{bloomfield_characterizing_2020, giuntoli_seasonal_2022}, canonical correlation analysis (CCA) combined with k-means as an established statistical method to relate two high-dimensional input spaces, and R-VAE (regression variational autoencoder) combined with k-means clustering \cite{zhao_variational_2019-1} which is a precursor of the RMM-VAE method.

Overall, we find that the novel RMM-VAE method is able to improve the predictive skill of the identified probabilistic weather regimes with respect to the target variable while maintaining higher regime robustness and persistence compared to the other targeted methods. The method thereby balances the different objectives of targeted clustering well, and better than the other methods assessed in this study.

Evaluating the identified weather regimes, we find that all the targeted methods resolve the dynamical drivers of extreme precipitation better than the non-targeted baseline PCA + k-means approach (Figure \ref{cluster_centers}). Moreover, we find that the R-VAE method performs best in predicting the target variable from the weather regimes assignment, followed by RMM-VAE up to a certain cluster number. Investigating the persistence and separability of the regimes as proxies for their physical robustness, we find that while all targeted methods perform worse than PCA, the RMM-VAE method performs best among the targeted methods (Figure \ref{persistence}).

By investigating the dimensionality-reduced spaces estimated by the different methods (Figure \ref{tsne}), we find that the two VAE methods disentangle the dimension of the geopotential height field associated with variations in precipitation over Morocco. However, while the R-VAE + k-means method subsequently clusters geopotential height data in bands along this disentangled dimension, integrating the probabilistic clustering with the targeted dimensionality reduction in the RMM-VAE method appears to identify more coherent clusters. 

Analysing the reconstruction of the input space from the reduced representations, we find that both VAE methods outperform the two linear methods (PCA + k-means and CCA + k-means) in terms of reconstruction loss (Figure \ref{rmse_boxplot}). While a more efficient reconstruction of the input space is expected from a generic VAE architecture due to the possibility of fitting a non-linear encoding function \citep{murphy_probabilistic_2023}, this result is not obvious here, given that the presented RMM-VAE method has the two additional objectives of disentangling a scalar target variable and fitting probabilistic clusters. 

The results highlight two trade-offs in identifying weather regimes targeted to a local-scale impact variable. First, regimes that are more predictive of an impact variable risk compromising their physical robustness, assessed here through regime persistence and separability (compare Figure \ref{rpss} and Figure \ref{persistence}). Although the R-VAE + k-means method identifies the most targeted regimes, the method performs worst in terms of cluster persistence and separability. This loss of physical robustness can imply that the predictability of the regimes themselves is reduced which is undesirable for their use in applications such as extended-range forecasting. This trade-off was encountered by \cite{bloomfield_patternbased_2021} who found that clustering their target variable directly maximized information about the impact \citep{bloomfield_characterizing_2020} but came at the cost of significantly reduced regime predictability \citep{bloomfield_patternbased_2021}. The second trade-off is that more predictive clusters can be achieved by focusing the dimensionality reduction on a subset of the input space, as we see in Figure \ref{rmse_spatial} in the case of CCA, as well as the two VAE methods if the reconstruction loss is down-weighted using the $\beta$ parameter. However, the resulting regimes do not contain information on the full atmospheric phase space which implies a loss of information on transition dynamics. This trade-off is implicitly encountered in methods that filter input days to the occurrence of extremes, thereby manually subsetting the input space \citep{rouges_european_2023}.

The RMM-VAE method performs well in navigating these trade-offs. Among the targeted clustering methods, it identifies the most robust clusters, while performing second-best only to the other VAE method in terms of predictive skill. In contrast, the CCA + k-means method identifies clusters that attain a similarly low separability score as the R-VAE + k-means method, while achieving a significantly lower predictive skill. This result highlights the benefit of using a probabilistic machine learning method to fit a non-linear dimensionality reduction that also allows incorporating a prediction target in a statistically coherent model. Furthermore, the RMM-VAE method makes the trade-offs involved in identifying targeted weather regimes explicit and allows their expression as part of the loss function of the model (Equation \ref{rmmvae_loss}), rather than addressing them implicitly through the choice of cluster number \citep{gadouali_link_2020} or pre-filtering \citep{rouges_european_2023}. In addition, the RMM-VAE method identifies probabilistic weather regimes that can give valuable information on transitional states, to be investigated in future work.

The proposed RMM-VAE method has limitations that should be addressed in future work. For example, we find that by using the full precipitation field as input, CCA is able to identify different spatial patterns of extreme precipitation associated with different regimes. This stands in contrast to the VAE methods that are trained using total precipitation over Morocco as a scalar target variable, and therefore primarily disentangle the dynamical drivers of rainfall over regions along the coast that contribute more to total rainfall. To address this, the RMM-VAE method could be further developed to incorporate higher-dimensional target variables.

The RMM-VAE method could also be applied to other regions and target variables, including more realistic and decision-relevant impact variables. In particular, precipitation based on ERA5 reanalysis data, the target variable used here, has known biases, especially in the pre-satellite era \citep{lavers_evaluation_2022}. While the impact variable explored here is 3-day precipitation over Morocco, any impact variable that has a justifiable link to the large-scale meteorological variables, such as renewable energy supply or the number of people impacted by an extreme event, could be used. Furthermore, the sensitivity of the regimes to pre-processing steps such as the choice of geographical region and the lowpass filter applied to the data could also be further investigated. Future work could also assess the predictability of the regimes themselves, their decadal variability, as well as their relationship to known teleconnections. This would improve the understanding of their usefulness for applications such as sub-seasonal to seasonal forecasting, dynamical adjustment and statistical downscaling of climate models.

The RMM-VAE method presented in this paper contributes a novel probabilistic machine learning method to statistically relate large-scale atmospheric dynamics to regional extremes in local-scale impact variables. The method shows promise in identifying weather regimes that disentangle the dynamical drivers of the target variable while maintaining the physical robustness of the regimes better than other methods, indicating its potential usefulness for a range of climate applications. The results also give further insight into the trade-offs involved in targeting weather regimes to a local impact variable. Overall, this contribution aims to highlight the benefits of motivating and guiding method development in machine learning with a physical research question and understanding of atmospheric dynamics, hopefully contributing to the further development of suitable machine learning methods in this field.

\paragraph{Acknowledgments}
The authors thank Jakob Wessel for useful discussions and feedback. ERA5 reanalysis data \citep{hersbach_era5_2020} was downloaded from the Copernicus Climate Change Service (C3S) (2023). The results contain modified Copernicus Climate Change Service information 2020. Neither the European Commission nor ECMWF is responsible for any use that may be made of the Copernicus information or data it contains.

\paragraph{Funding Statement}
TGS and MK were funded by the European Commission Horizon 2020 project XAIDA (Extreme Events: Artificial Intelligence for Detection and Attribution), Grant Agreement No. 101003469. FRS is funded by the Advancing the Frontiers of Earth System Prediction (AFESP) Doctoral Training Programme.

\paragraph{Competing Interests}
The authors declare no competing interests exist.

\paragraph{Data Availability Statement}
The data used in this study is available on Zenodo \url{https://zenodo.org/records/10101006}. The scripts necessary to reproduce the results in this study can be found on GitHub: \url{https://github.com/fiona511/RMM-VAE}.

\paragraph{Ethical Standards}
The research meets all ethical guidelines, including adherence to the legal requirements of the study country.

\paragraph{Author Contributions}
 Conceptualization: F.S., M.K., T.S., Y.K., M.B. 
 Methodology: F.S., M.K., Y.K.
 Investigation, software, data curation: F.S.
 Visualisation: F.S., M.K.
 Supervision: M.K., T.S.
 Writing original draft: F.S., M.K., T.S;
 Writing review and editing: F.S., M.K., T.S., Y.K., M.B.
 All authors approved the final submitted draft.

\bibliographystyle{plainnat}
\bibliography{references}

\newpage

\section{Appendix A - loss function derivations for R-VAE and RMM-VAE}\label{appendixloss}

\subsection{R-VAE}

\begin{align}
\mathcal{L}(\vect{x}) &= -D_{KL} \left(q_\phi(\vect{z}, t \mid \vect{x}) \mid p_\theta(\vect{z}, \vect{x}, t) \right) \\
&= - \int_{z, t} q_\phi (\vect{z} \mid \vect{x}) q_\phi (t \mid \vect{x}) \log \frac{q_\phi (\vect{z} \mid \vect{x}) q_\phi (t \mid \vect{x})}{p_\theta(\vect{x} \mid \vect{z}) p_\theta(\vect{z} \mid t) p_\theta(t)} \nonumber \\
&= \int_z q_\phi (\vect{z} \mid \vect{x}) \log p_\theta(\vect{x} \mid \vect{z}) - \int_{z,t} q_\phi (\vect{z} \mid \vect{x}) q_\phi (t \mid \vect{x}) \log \frac{q_\phi (\vect{z} \mid \vect{x})}{p_\theta(\vect{z} \mid t)} - \int_t q_\phi (t \mid \vect{x}) \log \frac{q_\phi (t \mid \vect{x})}{p_\theta (t)} \nonumber \\
&= \mathbb{E}_{q_\phi(\vect{z} \mid \vect{x})}\left[ \log p_\theta(\vect{x} \mid \vect{z}) \right] - \mathbb{E}_{q_\phi(t \mid \vect{x})}\left[D_{KL}(q_\phi(\vect{z} \mid \vect{x}) \mid p_\theta(\vect{z} \mid t))  \right] - D_{KL} \left(q_\phi (t \mid \vect{x}) \mid p_\theta (t) \right). \nonumber
\end{align}

\subsection{RMM-VAE}

\begin{align}
\mathcal{L}(\vect{x}) &= -D_{KL}(q_\phi (\vect{z},\vect{c},t \mid \vect{x}) \mid  p_\theta (\vect{x}, \vect{z}, t, \vect{c})) \nonumber \\
&= - \sum_k \int_{z,t} q_\phi (\vect{z} \mid \vect{x}) q_\phi (t \mid \vect{x}) q_\phi (\vect{c^k} \mid \vect{x}) \log \frac{q_\phi (\vect{z} \mid \vect{x}) q_\phi (t \mid \vect{x}) q_\phi (\vect{c^k} \mid \vect{x})}{p_\theta(\vect{x} \mid \vect{z}) p_\theta(\vect{z} \mid t) p_\theta(\vect{z} \mid \vect{c^k}) p(\vect{c^k})} \nonumber \\
&= \sum_k q_\phi (\vect{c^k} \mid \vect{x}) \bigg[ \int_z q_\phi (\vect{z} \mid \vect{x}) \log p_\theta(\vect{x} \mid \vect{z}) - \int_z q_\phi (\vect{z} \mid \vect{x}) \log p_\theta(\vect{x} \mid \vect{z}) \nonumber \\
&- \int_{z,t} q_\phi (\vect{z} \mid \vect{x}) q_\phi (t \mid \vect{x}) \log \frac{q_\phi (\vect{z} \mid \vect{x})}{p_\theta(\vect{z}\mid t)} \nonumber \\
 &- \int_t q_\phi (t \mid \vect{x}) \log \frac{q_\phi (t \mid \vect{x})}{p(t)} - \int_z q_\phi (\vect{z} \mid \vect{x}) \frac{q_\phi (\vect{z} \mid \vect{x})}{p_\theta(\vect{z} \mid \vect{c^k})} - \log \frac{q_\phi (\vect{c^k} \mid \vect{x})}{p(\vect{c^k}} \bigg] \nonumber \\
& = \sum_k q_\phi (\vect{c^k} \mid \vect{x}) \bigg[\mathbb{E}_{q_\phi (\vect{z} \mid \vect{x})} [\log p_\theta(\vect{x} \mid \vect{z})] - \mathbb{E}_{q_\phi (t \mid \vect{x})} [D_{KL} (q_\phi (\vect{z} \mid \vect{x}) \mid p(\vect{z} \mid t))] \nonumber \\
&- D_{KL} (q_\phi(t \mid \vect{x}) \mid p(t)) 
 - D_{KL} (q_\phi (\vect{z} \mid \vect{x}) \mid p_\theta(\vect{z} \mid \vect{c^k})) \bigg] -D_{KL}(q_\phi (\vect{c^k} \mid \vect{x}) \mid p(\vect{c^k})).
\end{align}

\section{Appendix B - sensitivity analysis}\label{appendixA}

To analyse the robustness of the results, the sensitivity of the results to two pre-processing steps is analysed: 
\begin{itemize}
\item Averaging z500 over 3 days vs 5 days: qualitatively similar results, very similar predictive skill, ordering of latent space and reconstruction error. 3-day clusters have a slightly decreased persistence and silhouette score, as expected.
\item Selecting a slightly larger geographical region (westward shift and overall larger region tried). This influences all evaluated metrics, however, the ranking of different methods in the different evaluation categories remains the same.
\end{itemize}
So while these choices do affect the precise values of different evaluation metrics, they do not affect the overall ordinal findings when comparing the methods.

\begin{figure}[!]%
\centering
\includegraphics[width=0.9\textwidth]{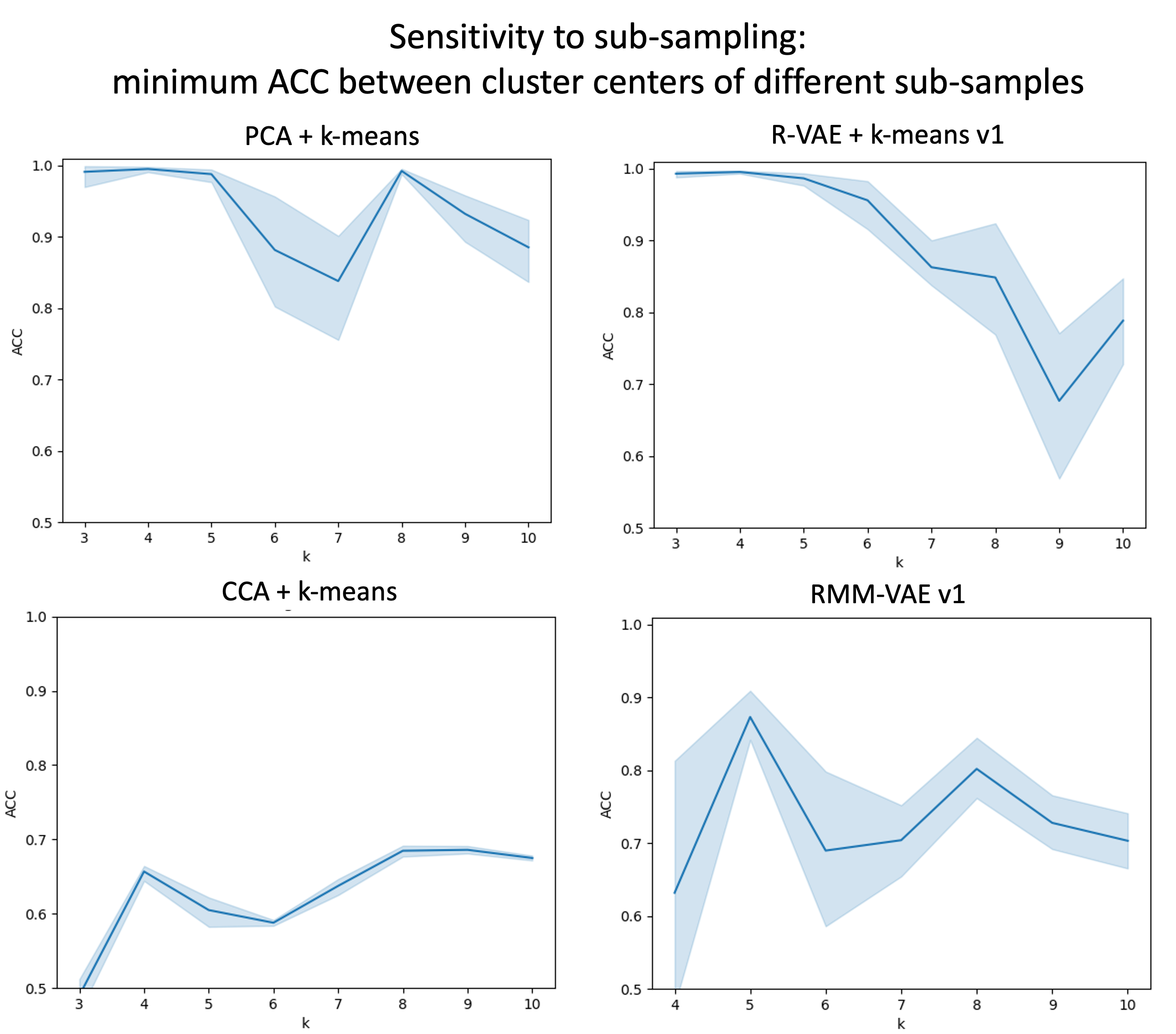}
{\caption{50 subsamples containing 80\% of data points each are created, and cluster centres are computed. The two sets of cluster centres are then paired by matching centres with the lowest Anomaly Correlation Coefficient (ACC). The lowest of these maximum ACC values is recorded, corresponding to the ACC of the least well-correlated cluster pair}
\label{acc}}
\end{figure}

Furthermore, the sensitivity of cluster centres to subsampling the data is analyzed. This is important because it is not desirable for the cluster centres to change drastically with new data. Figure \ref{acc} shows the anomaly correlation coefficient (ACC) between the cluster centres computed from subsamples of the data and the cluster centres computed from the full data. We find that the sensitivity of the cluster centres to subsampling depends on the method and cluster number. Overall, the PCA + k-means and R-VAE + k-means methods produce very robust clusters (ACC>0.95) to subsampling up to a cluster number of 5. Beyond $k=5$ the robustness to subsampling degrades consistently for R-VAE + k-means, while it peaks again at $k=8$ for PCA + k-means. Although the robustness of cluster centres is overall slightly lower for the RMM-VAE method, the most robust cluster numbers are consistent with the PCA + k-means method. The cluster centres identified by the CCA + k-means methods are the least robust to subsampling, with an ACC smaller than 0.7 for all cluster numbers analysed. This analysis justifies illustrating the resulting cluster centres for $k=5$ in the results section and highlights that the choice of $k$ depends on the method chosen.

\section{Appendix C - sample cluster timeseries}\label{appendixB}

\begin{figure}[!]%
\includegraphics[width=0.9\textwidth]{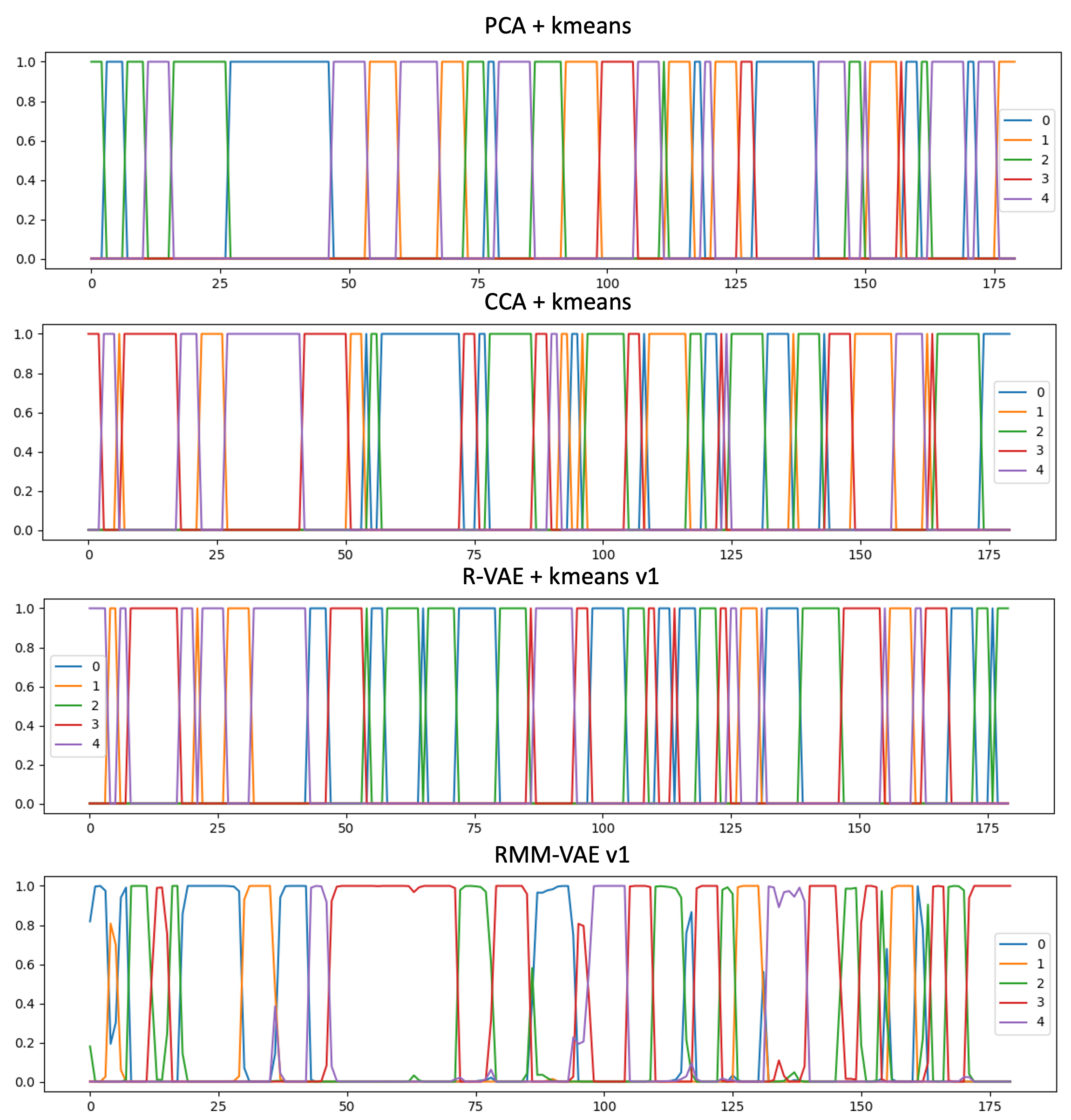}
{\caption{Sample time series of cluster assignment in different methods.}
\label{timeseries}}
\end{figure}

\section{Appendix D - Statistical glossary}\label{appendixC}

\begin{table}[!]
\tabcolsep=0pt%
\caption{Glossary of selected statistical and machine learning terminology based on \cite{murphy_probabilistic_2022} \label{glossary}}
\begin{tabular*}{\textwidth}{ p{0.25\textwidth}  p{0.75\textwidth} }\toprule
Backpropagation & Algorithm to calculate the gradient of a loss function and implement gradient descent used to train neural networks. Part of the two-step cycle used to train neural networks: during the forward pass, the output of the neural network is computed given the current weights of the network. During the backward pass or backpropagation, the weights of the hidden layers of the network are adjusted to reduce the loss function. \\\midrule
Divergence & Distance metric between two probability distributions $p$ and $q$, required to satisfy $D(p, q) \geq 0$ with equality iff $p=q$, symmetry and triangle inequality. The Kullback-Leibler divergence can be interpreted as the information lost by representing $p$ with $q$, and is defined as the difference between the negative entropy of $p$ and the crossentropy between $p$ and $q$ \citep{kullback_information_1951}. \\\midrule
Latent space & Also known as embedding space. A d-dimensional vector space that encodes information about a higher-dimensional space in a meaningful representation. \\\midrule
Prior probability & In Bayesian statistics, the prior probability refers to the assumed probability of a variable before selected data or evidence are considered. \\\midrule
Posterior probability & In Bayesian statistics, the posterior probability corresponds to the conditional probability of statistical model parameters given the likelihood of the data and the selected prior. \\\midrule
Regularization & Penalty term added when fitting a statistical model used to prevent overfitting. \\\midrule
\end{tabular*}
\end{table}

\end{document}